\input harvmac.tex 

\lref\batvil{
I.~A.~Batalin and G.~A.~Vilkovisky,
%``Quantization Of Gauge Theories With Linearly Dependent Generators,''
Phys.\ Rev.\ D {\bf 28}, 2567 (1983)
[Erratum-ibid.\ D {\bf 30}, 508 (1984)].
%%CITATION = PHRVA,D28,2567;%%
}

\lref\charpress{ V. Chari and A. Pressley,
 ``A guide to quantum groups, '' CUP 1994  }

\lref\maj{ S. Majid, ``Foundations of quantum group theory'' CUP 1995 }

\lref\cds{
A.~Connes, M.~R.~Douglas and A.~Schwarz,
%``Noncommutative geometry and matrix theory: Compactification on tori,''
JHEP {\bf 9802}, 003 (1998)
[arXiv:hep-th/9711162].
%%CITATION = HEP-TH 9711162;%%
}

\lref\lnr{
D.~A.~Lowe, H.~Nastase and S.~Ramgoolam,
%``Massive IIA string theory and Matrix theory compactification,''
Nucl.\ Phys.\ B {\bf 667}, 55 (2003)
[arXiv:hep-th/0303173].
%%CITATION = HEP-TH 0303173;%%
}

\lref\aikkt{
H.~Aoki, N.~Ishibashi, S.~Iso, H.~Kawai, Y.~Kitazawa and T.~Tada,
%``Noncommutative Yang-Mills in IIB matrix model,''
Nucl.\ Phys.\ B {\bf 565}, 176 (2000)
[arXiv:hep-th/9908141].
%%CITATION = HEP-TH 9908141;%%
}

\lref\kimura{
Y.~Kimura,
%``Noncommutative gauge theory on fuzzy four-sphere and matrix model,''
Nucl.\ Phys.\ B {\bf 637}, 177 (2002)
[arXiv:hep-th/0204256].
%%CITATION = HEP-TH 0204256;%%
}

\lref\branho{
R.~Brandenberger and P.~M.~Ho,
%``Noncommutative spacetime, stringy spacetime uncertainty principle, and  density fluctuations,''
Phys.\ Rev.\ D {\bf 66}, 023517 (2002)
[AAPPS Bull.\  {\bf 12N1}, 10 (2002)]
[arXiv:hep-th/0203119].
%%CITATION = HEP-TH 0203119;%%
}

\lref\goodwal{ Goodman and Wallach, ``Representations and 
Invariants of classical Groups,'' section 5.2.3 } 
\lref\iktw{ 
S.~Iso, Y.~Kimura, K.~Tanaka and K.~Wakatsuki,
%``Noncommutative gauge theory on fuzzy sphere from matrix model,''
Nucl.\ Phys.\ B {\bf 604}, 121 (2001)
[arXiv:hep-th/0101102].
%%CITATION = HEP-TH 0101102;%%
}
\lref\grrefi{ I.S.Gradshtein and I.M.Rhyzik, Academic Press 2000,
  Eq. 0.160.2 }
\lref\grrefii{  I.S.Gradshtein and I.M.Rhyzik, Academic Press 2000,
  Eq. 0.156.1 }

\lref\sphdiv{ S. Ramgoolam, ``On spherical harmonics for 
fuzzy spheres  in diverse dimensions,''  hep-th/0105006 } 
\lref\vilen{ Vilenkin book  } 

\lref\gkp{ H. Grosse, C. Klimcik and P. Presnajder,
``On finite  4-D Quantum Field Theory in noncommutative geometry,''
[hep-th/9602115],  Commun.Math.Phys.180:429-438,1996.
%%CITATION = HEP-TH 9602115;%%
}

\lref\hullw{ 
C.~M.~Hull,
%``Lectures on W gravity, W geometry and W strings,''
arXiv:hep-th/9302110.
%%CITATION = HEP-TH 9302110;%%
}

\lref\sphdiv{ S.~Ramgoolam,
``On spherical harmonics for fuzzy spheres in diverse dimensions,''
Nucl.\ Phys.\ B {\bf 610}, 461 (2001)
[hep-th/0105006].
%%CITATION = HEP-TH 0105006;%%
}

\lref\clt{J. Castellino, S. Lee, W. Taylor,
``Longitudinal 5-branes as 4-spheres in matrix theory,'' 
[hep-th/9712105], Nucl.Phys. B526 (1998) 334-350.
%%CITATION = HEP-TH 9712105; %% 
} 

\lref\cmt{ N. Constable, R. Myers and O. Tafjord,
``Non-abelian brane intersections,''
[hep-th/0102080],  JHEP 0106:023,2001.
%%CITATION = HEP-TH 0102080; %%
}

\lref\kramer{ M. Kramer, `` Some remarks suggesting an interesting 
 theory of harmonic functions on $SU(2n+1)/Sp(n)$ and 
$SO(2n+1)/U(n)$,'' Arch. Math. 33 ( 1979/80), 76-79. }

\lref\iktw{ S.~Iso, Y.~Kimura, K.~Tanaka and K.~Wakatsuki,
``Noncommutative gauge theory on fuzzy sphere from matrix model,''
Nucl.\ Phys.\ B {\bf 604}, 121 (2001)
[hep-th/0101102].
%%CITATION = HEP-TH 0101102;%%
}

\lref\gelf{ I.M.Gelfand, A.V. Zelevinskii,
``Models of representations
of classical groups and their hidden symmetries,'' 
Funct. An. Priloz. vol. 18, 1984. } 

\lref\zr{ Z. Guralnik,  S. Ramgoolam, 
``On the polarization of unstable D0-branes
 into non-commutative odd spheres,''
[hep-th/0101001], JHEP 0102 (2001) 032.
%%CITATION = HEP-TH 0101001;%%
}

\lref\ho{ P.~M.~Ho,
``Fuzzy sphere from matrix model,''
[hep-th/0010165], JHEP {\bf 0012}, 015 (2000).
%%CITATION = HEP-TH 0010165;%%
}
\lref\jr{ A. Jevicki and S. Ramgoolam,
`` Noncommutative gravity from the ADS/CFT correspondence,''
[hep-th/9902059], JHEP {\bf 9904} (1999) 032.
%%CITATION = HEP-TH 9902059;%%
 }

\lref\gutze{F. Gursey and C.H. Tze, ``On the role of division,
Jordan and related algebras in particle physics,''
World Scientific 1996.}

\lref\wang{H. C. Wang,
Nagoya. Math. J. {\bf 13}, 1 (1958).}

\lref\Myers{ R. C. Myers,
``Dielectric-branes'',JHEP 9912 (1999) 022
[hep-th/9910053].
%%CITATION = HEP-TH 9910053;%%
}

\lref\fulhar{ W. Fulton and G. Harris,
``Representation theory,'' 
Springer Verlag 1991.} 

\lref\BFSS{ ``T. Banks, N. Seiberg and S. Shenker, 
M-Theory as a matrix model : a conjecture,''
[hep-th/9610043],  Phys.Rev.D55:5112-5128,1997.
%%CITATION = HEP-TH 9610043;%%
 } 

\lref\hrt{ P.M.Ho, S.Ramgoolam and R.Tatar,
``Quantum space-times and finite N
effects in 4-D superyang-mills theories,''
[hep-th/9907145], 
Nucl.Phys.{\bf B573} (2000) 364,
%%CITATION = HEP-TH 9907145
}

\lref\dwhn{  B. de Wit, J. Hoppe, H. Nicolai, 
``On the quantum mechanics of supermembranes,''
Nucl.Phys.B305:545,1988.
%%CITATION = NUPHA,B305,545;%%
}

\lref\madore{ J. Madore, ``The Fuzzy sphere, ''
Class.Quant.Grav.9:69-88,1992  
%%CITATION = CQGRD,9,69;%% 
}

\lref\cmm{
P.~Cook, R.~de Mello Koch and J.~Murugan,
%``Non-Abelian bionic brane intersections,''
arXiv:hep-th/0306250.
%%CITATION = HEP-TH 0306250;%%
}

\lref\berkverl{ M. Berkooz and H. Verlinde,
``Matrix theory, AdS/CFT and
Higgs-Coulomb equivalence,''
[hep-th/9907100], 
JHEP 9911 (1999) 037.
%%CITATION = HEP-TH 9907100;%%
} 

\lref\bss{ T. Banks, N. Seiberg, S. Shenker, `` Branes form
Matrices,'' [hep-th/9612157], \hfill\break
Nucl.Phys. B490:91-106,1997 
%%CITATION = HEP-TH 9612157;%% 
}

\lref\grt{ O. Ganor, S. Ramgoolam, W. Taylor,
``Branes, fluxes and duality in M(atrix)-theory,''
[hep-th/9611202], 
 Nucl.Phys.B492:191-204,1997 
%%CITATION = HEP-TH 9611202;%% 
}

\lref\AH{ N.~Arkani-Hamed, A.~G.~Cohen and H.~Georgi,
``(De)constructing dimensions,''
[hep-th/0104005],
Phys.\ Rev.\ Lett.\  {\bf 86}, 4757 (2001).
%%CITATION = HEP-TH 0104005;%%
}

\lref\bdlmo{ 
A.P. Balachandran, B. P. Dolan, J.H. Lee, X. Martin, D. O'Connor
``Fuzzy complex projective spaces and their star products,''
[hep-th/0107099]
%%CITATION = HEP-TH 0107099;%%
 }

\lref\tv{
S. P. Trivedi, S. Vaidya,
``Fuzzy cosets and their gravity duals,''
[hep-th/0007011], JHEP 0009:041,2000 
%%CITATION = HEP-TH 0007011;%%
}

\lref\connes{ A. Connes, ``Non-commutative geometry,'' Academic Press,
 1994.  }   

\lref\ps{ P.J.Silva,`` Matrix string theory and the Myers effect,''
[hep-th/0111121].
%%CITATION = HEP-TH 0111121;%%
}

\lref\rief{ M.A.Rieffel
``Matrix algebras converge to the sphere 
for quantum Gromov--Hausdorff distance,'' [math.OA/0108005].
%%CITATION = MATH-OA 0108005;%% 
}

\lref\hdim{
P.~M.~Ho and S.~Ramgoolam,
``Higher dimensional geometries from matrix brane constructions,''
Nucl.\ Phys.\ B {\bf 627}, 266 (2002)
[arXiv:hep-th/0111278].
%%CITATION = HEP-TH 0111278;%%
}

\lref\aiikkt{
H.~Aoki, N.~Ishibashi, S.~Iso, H.~Kawai, Y.~Kitazawa and T.~Tada,
%``Noncommutative Yang-Mills in IIB matrix model,''
Nucl.\ Phys.\ B {\bf 565}, 176 (2000)
[arXiv:hep-th/9908141].
%%CITATION = HEP-TH 9908141;%%
}

\lref\IKKT{
N.~Ishibashi, H.~Kawai, Y.~Kitazawa and A.~Tsuchiya,
%``A large-N reduced model as superstring,''
Nucl.\ Phys.\ B {\bf 498}, 467 (1997)
[arXiv:hep-th/9612115].
%%CITATION = HEP-TH 9612115;%%
}

\lref\seib{ 
N.~Seiberg,
%``A note on background independence in noncommutative gauge theories, 
matrix model and tachyon condensation,''
JHEP {\bf 0009}, 003 (2000)
[arXiv:hep-th/0008013].
%%CITATION = HEP-TH 0008013;%%
}

\lref\ars{
A.~Y.~Alekseev, A.~Recknagel and V.~Schomerus,
%``Non-commutative world-volume geometries: Branes on SU(2)
% and fuzzy  spheres,''
JHEP {\bf 9909}, 023 (1999)
[arXiv:hep-th/9908040].
%%CITATION = HEP-TH 9908040;%%
}
\lref\hidclass{
A.~Jevicki, M.~Mihailescu and S.~Ramgoolam,
%``Hidden classical symmetry in quantum spaces at roots of unity : 
%From  q-sphere to fuzzy sphere,''
arXiv:hep-th/0008186.
%%CITATION = HEP-TH 0008186;%%
}

\lref\gms{
H.~Grosse, J.~Madore and H.~Steinacker,
%``Field theory on the q-deformed fuzzy sphere. II: Quantization,''
J.\ Geom.\ Phys.\  {\bf 43}, 205 (2002)
[arXiv:hep-th/0103164].
%%CITATION = HEP-TH 0103164;%%
}

\lref\cgz{
T.~L.~Curtright, G.~I.~Ghandour and C.~K.~Zachos,
%``Quantum Algebra Deforming Maps, Clebsch-Gordan Coefficients, 
Coproducts, U And R Matrices,''
J.\ Math.\ Phys.\  {\bf 32}, 676 (1991).
%%CITATION = JMAPA,32,676;%%
}

\lref\pmnass{
P.~M.~Ho,
%``Making non-associative algebra associative,''
JHEP {\bf 0111}, 026 (2001)
[arXiv:hep-th/0103024].
%%CITATION = HEP-TH 0103024;%%
}
  
\lref\cornschiap{
 L.~Cornalba and R.~Schiappa,
%``Nonassociative star product deformations for D-brane worldvolumes in  curved backgrounds,''
Commun.\ Math.\ Phys.\  {\bf 225}, 33 (2002)
[arXiv:hep-th/0101219].
%%CITATION = HEP-TH 0101219;%%
}

\lref\drinf{ 
V.G. Drinfeld, { \it Leningrad Math. Jour.} { \bf 1 }, 1419, 1990 } 

\lref\bchtz{ B.A.Bernevig, C-H Chern, J-P Hu, N. Toumbas, S-C Zhang, 
cond-mat/0206164,  Annals Phys. 300 (2002) 185 } 

\lref\zhhu{ S-C Zhang, J-P Hu
``A Four Dimensional Generalization of the Quantum Hall Effect,''
Science 294 (2001) 823 } 

\lref\smaj{
 E.~Batista and S.~Majid,
%``Noncommutative geometry of angular momentum space U(su(2)),''
J.\ Math.\ Phys.\  {\bf 44}, 107 (2003)
[arXiv:hep-th/0205128].
%%CITATION = HEP-TH 0205128;%%
}

\lref\sjab{
A.~B.~Hammou, M.~Lagraa and M.~M.~Sheikh-Jabbari,
%``Coherent state induced star-product on R(lambda)**3 and the fuzzy  sphere,''
Phys.\ Rev.\ D {\bf 66}, 025025 (2002)
[arXiv:hep-th/0110291].
%%CITATION = HEP-TH 0110291;%%
}

\lref\seibwit{ 
N.~Seiberg and E.~Witten,
%``String theory and noncommutative geometry,''
JHEP {\bf 9909}, 032 (1999)
[arXiv:hep-th/9908142].
%%CITATION = HEP-TH 9908142;%%
}

\vskip 1cm

 \Title{ \vbox{\baselineskip12pt\hbox{  Brown-Het-1368  }}
 \vbox{\baselineskip12pt\hbox{  QMUL-PH-0310  }}
}
 {\vbox{
\centerline{  Towards Gauge theory for a class of }
\centerline{   commutative and non-associative  fuzzy spaces   }  }}

\centerline{ { Sanjaye Ramgoolam }}
\centerline{{\sl Queen Mary College }}
\centerline{{\sl London, E1 4NS }}
\smallskip
\centerline{{\tt  s.ramgoolam@qmul.ac.uk }}
\vskip .3in

% We consider  derivatives obeying a deformed Leibniz rule, acting on 
% commutative but non-associative algebras 
% related to the $SO(2k+1)$ covariant fuzzy $2k$-sphere ( with $k$ 
% larger than 2 ), where the product is defined by a
% projection of the  matrix product. 
% In the limit of large matrices ( large $n$), these derivatives 
% approach ordinary derivatives of classical geometry. There is  
% a star product, related to the projected Matrix product by a twist,
% for which the derivatives obey the standard Leibniz rule  at finite $n$. 
% These ingredients are used to 
% construct  gauge theories  on the commutative non-associative algebra
% of a fuzzy $2k$-sphere and a related deformed algebra of polynomial 
% functions on $R^{2k+1}$.  A consequence of non-associativity is that gauge
% fields and gauge parameters have to be functions of coordinates as well as
% derivatives. The usual gauge fields depending on coordinates only 
% are recovered after a partial gauge fixing. The  action of a deformed  
% $ SO(2k+1,1)$ conformal algebra naturally appears in the finite $n$ 
% fuzzy $2k$-sphere. The deformation parameter for these commutative but 
% non-associative algebras is a scalar of the rotation group, and suggest 
% interesting string-inspired algebraic deformations of spacetime 
% which preserve Lorentz-invariance. 

  We discuss gauge theories for commutative but non-associative 
  algebras related to the $ SO(2k+1)$ covariant finite dimensional 
  fuzzy $2k$-sphere algebras.  A consequence of non-associativity is that gauge
  fields and gauge parameters have to be generalized to be 
  functions of coordinates as well as derivatives. The usual 
  gauge fields depending on coordinates only 
  are recovered after a partial gauge fixing.The deformation parameter
  for these commutative but non-associative algebras is a scalar
  of the rotation group. This suggests interesting string-inspired  algebraic
  deformations of spacetime  which preserve Lorentz-invariance.

\Date{ October  2003 }

\def\Ga{ \Gamma } 
 
\def\cRn{ {\cal{R}}_n }

\def\cAn{ {\cal{A}}_n }   
 
\def\cBn{  { {\cal{B}}_n }   } 
\def\cBnst{ { {\cal{B}}_n^*  }}
\def\cBnc{  { {\cal{B}}^c_n }}

\def\tlddel#1#2#3{ { ~~ \tilde 
\delta (~  {#1} ( {#2} ) ~,~ {#1}
( {#3} ) ~ ) ~ } } 
\def\dr#1{  { \partial \over \partial z_{ #1 } }}
  
\def\prtder#1{ { \partial \over \partial  #1  }}

%\draftmode 

\newsec{ Introduction } 

  The fuzzy four-sphere  \gkp\ has several applications in 
  D-brane physics \clt\cmt. While the fuzzy four-sphere 
 has some similarities to the fuzzy two-sphere \madore, 
 it is also different in important ways. The Matrix Algebra 
 of the fuzzy $2k$-sphere at finite $n$ is the space of
  transformations, $ End ( \cRn )$,  of an irreducible representation,  
 $ \cRn $,  of $Spin (2k+1)$. 
 It  contains a truncated  spectrum of symmetric traceless 
  representations  of  $SO(2k+1)$ which form a truncated 
 spectrum of spherical harmonics, but contains additional 
 representations. The analysis of the representations 
 in the Matrix Algebras for fuzzy spheres in diverse dimensions 
 was given in \sphdiv.  For even spheres 
 $ S^{2k}$ these representations span, at large $n$,  the 
 space of functions on a higher dimensional coset 
 $SO(2k+1)/U(k)$, which are bundles over the 
 spheres $S^{2k}$.  The extra degrees of freedom 
 in the Matrix algebra  can be interpreted  equivalently 
 in terms of non-abelian fields on the spheres \hdim.
 In the case of the fuzzy 
 four sphere the higher dimensional geometry is 
 $SO(5)/U(2)$ which is one realization of $CP^3$ as a 
 coset. 
 
 Any discussion of field theory on the fuzzy  sphere $S^{2k}$ 
 requires a product for the fields. The configuration space of a
 scalar field  on the fuzzy four-sphere, is the subspace of $ End ( \cRn )$
 which transforms in the traceless symmetric representations.  
 This vector subspace of $ End ( \cRn )$ admits 
 an obvious product. It is obtained by taking  the Matrix product
 followed by a projection  back to the space of Matrices transforming as 
 symmetric traceless
 representations. The vector space of truncated spherical harmonics 
 equipped with this product 
 is denoted by $ \cAn ( S^{2k} )$. The product is
 denoted by $m_2$, which can be viewed as a map from 
 $ \cAn ( S^{2k} ) \otimes \cAn ( S^{2k} )$ to $ \cAn ( S^{2k} ) $.
 It is important 
 to distinguish  $ End ( \cRn ) $
 and $ \cAn ( S^{2k} )$,  which are the 
 full Matrix algebra and the algebra obtained 
 after projection, respectively. The product 
 on $ \cAn$ is non-associative and commutative 
 but the non-associativity vanishes at large $n$ \sphdiv\hdim. 
 The Matrix algebra $  End ( \cRn )   $ contains 
 Matrices $ X_{\mu \nu } $  transforming in the adjoint of $SO(2k+1)$. 
 It also contains Matrices  $ X_{\mu } $ transforming  in the vector of 
 $SO(2k+1)$. The vector of $SO(2k+1) $ and the adjoint 
 combine into the adjoint of $SO(2k+2)$. The $SO(2k+2)$ acts 
 by commutators on the whole Matrix algebra. 
 
The projection used in defining the non-associative 
 product on $ \cAn (  S^{2k} )  $ commutes with 
 $SO(2k+1)$ but does not commute with $SO(2k+2)$. Hence 
 the generators of $SO(2k+1)$ provide derivations 
 on the $ \cAn (  S^{2k} )  $ \hdim. 
However  derivations transforming in the vector 
 of $SO(2k+1)$ would be very useful in developing 
 gauge theory for the $ \cAn ( S^{2k} )$, where 
 we would write a covariant derivative of the 
 form $ \delta_{ \alpha } - i A_{\alpha } $ 
 and use that as a building block for the gauge theory, 
 a technique that has found many applications in Matrix Theories
 \BFSS\IKKT  ( see for example \refs{ \grt,\cds, \aikkt, \seib , 
 \lnr,\iktw } ).

  The classical sphere can be  described in terms of an algebra
  generated by the coordinates $z_{\alpha}$ of the embedding $R^{2k+1}$, as
  a projection following from  the constraint $ \sum_{\alpha=1 }^{2k+1}
  z_{\alpha} z_{\alpha} = 1 $. When we drop the constraint of a
  constant radius, the algebra of functions on $R^5$ generated by the 
  $z_{\alpha}$ admits derivations $\partial_{\alpha}$ which obey
  $\partial_{\alpha} z_{\beta}  = \delta_{\alpha \beta } $.
  These translations can be projected to the tangent space of the sphere 
  to give derivations on the sphere.   The latter $P_{\alpha }$ 
 can be written as $ P_{\alpha} = \delta_{\alpha } - Q_{\alpha } $, 
 where $Q_{\alpha } $ is a  derivative  transverse to the sphere, 
 and can be characterized by its action on the Cartesian coordinates.

 The fuzzy  sphere algebra contains analogous operators $Z_{\alpha}$ 
 which obey a constraint  $\sum_{\alpha =1}^{2k+1} 
  Z_{ \alpha } Z_{\alpha } = { n+2k \over n  } $ 
 and form a finite dimensional algebra.  
 It is natural to consider at finite $n$, a  finite dimensional 
 algebra generated by the $ Z_{\alpha}$ obtained by  
 dropping this quadratic constraint,  but keeping the 
structure constants of the algebra.   This can be viewed as a 
 deformation of  the algebra of polynomial functions of 
 $Z_{\alpha}$ .  We call this finite dimensional algebra $ \cAn ( R^{2k+1}
  )$.  Imposing the constraints  
  $ \sum Z_{ \alpha } Z_{\alpha } = { n +2k \over n } $ on this deformed 
 polynomial algebra gives the algebra $ \cAn ( S^{2k})$. 
 Again one can consider `derivatives' $ \delta_{\alpha} $ defined such that 
 their action on the algebra $ \cAn ( R^{2k+1} )$ at large $n$ is just the 
 action of the generators of the translation group. In general they will not 
 satisfy the Leibniz rule at finite $n$.

 An approach to gauge theory on the 
 algebra  $ \cAn ( R^{2k+1} )$ is to study the deformation of 
 Leibniz Rule that is obeyed by $ \delta_{ \alpha } $. 
 This will allow us to obtain the gauge transformation 
 required of $ A_{ \alpha } $ such that the covariant derivative 
 $ D_{\alpha} = \delta_{\alpha} - i A_{\alpha } $ is indeed 
 covariant. We can expect the non-associativity to lead to 
 extra terms in the transformation of $A_{\alpha}$. 
 To get to gauge theory on $ \cAn (S^{2k+1} )$, we would define 
 the projection operators $ P_{ \alpha }, Q_{\alpha }$ 
 and study their deformed
 Leibniz rule, and then obtain the gauge transformation 
 rule for $A_{\alpha }$ by requiring that $P_{\alpha }  - i A_{\alpha}$
 is covariant. The strategy of doing non-commutative geometry 
 on the deformed $R^3$ as a way of getting at the non-commutative geometry 
 of the fuzzy $S^2$ is used in \refs{ \sjab , \smaj }. 

 The structure constants of the algebra $ \cAn ( S^{D} ) $
 and $  \cAn ( R^{D+1 } ) $ show an interesting  simplification at large $D$. 
 We will study  simpler algebras $ \cBn ( R^{2k+1} ) $  
 and $ \cBn ( S^{2k} ) $, which are obtained by 
 choosing $2k+1$ generators among the $D+1$ of $ \cAn ( R^{D+1} ) $. 
 By keeping $k$ fixed and letting $D$ be large we get 
 a simple algebra, whose structure constants can be easily 
 written down explicitly. This algebra is  still  
 commutative/non-associative, and allows us to explore 
 the issues of doing gauge theory in this 
 set-up. We explain the relation between $ \cBn ( S^{2k+1} ) $
 and $ \cAn ( S^{2k+1} ) $ in section 2.

 In section 3,  we explore  `derivatives' on $ \cBn ( R^{2k+1 } ) $ which obey 
 $ \delta_{\alpha} Z_{ \beta } = \delta_{ \alpha \beta }$. 
 To completely define these 
 operators we  specify  their action on a complete 
 basis of the deformed polynomial algebra. 
 The action we define  reduces in the large $n$ limit to the ordinary 
 action of infinitesimal translations. Having defined 
 the action of the derivatives  on the space of elements in the 
 algebra, we need to explore the interplay of the derivatives with the 
 product. The operation $ \delta_{ \alpha }$ 
 obeys the Leibniz rule at large $n$ but fails to do so 
 at finite $n$. The precise deformation of the Leibniz  rule 
 is described in section 3.

 In section 4  we show that by using a modified 
 product  $m_2^*$ ( which is also commutative and non-associative ) 
 the  derivatives $ \delta_{\alpha }$ do actually obey the 
 Leibniz rule. The modified product is related to the projected Matrix
 product $m_2$ by a `twisting'. While our use of the term `twist' 
 is partly colloquial, there is some similarity to the `Drinfeld  
 twist' that appears in quantum groups \cgz\drinf.
 Applications of the Drinfeld twist in recent literature
 related to fuzzy spheres include  
 \ars\hidclass\gms .

 In section 5,  using the $ \delta_{\alpha }$ and the product $m_2^*$ 
 we discuss an abelian gauge theory for the 
 finite deformed polynomial algebra $ \cBnst ( R^{2k+1}  )$. 
 The transformation law of $A_{\alpha}$,  which guarantees that 
 $ D_{ \alpha } $ is covariant, now has additional terms
 related to the associator. A detailed description of the associator 
 becomes necessary, some of which is given in  Appendix 2. 
 Appendix 1 is a useful prelude to the discussion of the 
associator. It gives the relation between the product 
$m_2^*$ and  the product $m_2^c$ which is
 essentially the classical product on polynomials. 
 The discussion of the gauge invariance shows 
 a key consequence of the non-associativity, that 
 gauge fields can pick up derivative dependences under a 
 gauge transformation by a gauge parameter which depends 
 only on coordinates. Hence we should enlarge our notion of 
 gauge field to include dependences on derivatives. This naturally 
 suggests that the gauge parameters should also be generalized to
 allow dependence on derivatives. We can then obtain ordinary 
 coordinate dependent gauge fields and  their gauge transformations 
 by a partial gauge fixing.  This discussion is somewhat formal 
 since a complete discussion of gauge theory on a deformation of 
 $R^{2k+1}$ should include a careful discussion of the integral. 
 We will postpone discussion of the integral for 
 the deformed $R^{2k+1}$ to the future. 
 However the structure of gauge fields, gauge parameters, gauge fixing 
 in the non-associative context uncovered here carries over 
 to the case of the gauge theory on $ \cBnst ( S^{2k} )$ which we discuss
 in the next section. One new ingredient needed here is 
 the projection of the translations of 
$R^{2k+1}$ to the tangent space of the sphere. This requires 
 introduction of new derivatives $ Q_{\alpha } $ such that 
 the projectors to the tangent space are $ P_{\alpha} =
 \delta_{\alpha} - Q_{\alpha}$.  The relevant properties of $ Q_{\alpha}$ 
 are described and the construction of abelian gauge theory on $ \cBnst ( S^{2k}
 )$ done in section 5. The discussion of the integral is much simpler
 in this case. Details on the deformed Leibniz rule for $Q_{\alpha}$ are 
 given in Appendix 3. 
%The final appendix is given for entertainment.
The final appendix 4, which is not used in the paper, 
but is an interesting technical extension of Appendix 1 
shows how to write the classical product $m_2^c$ in terms of $m_2$, the
product in $\cAn (R^{2k+1} )$   
  
 We end with conclusions and a discussion of avenues for 
 further research.

\newsec{ The Algebra  $\cBn  ( R^{2k+1} ) $  } 

We will define a deformed polynomial algebra 
$\cBn  ( R^{2k+1} ) $ 
starting from the fuzzy sphere algebras $ \cAn ( S^{D } ) $.
Consider operators in $ End ( \cRn ) $, where $ \cRn $ is 
the irrep of $Spin ( D+1 )$ whose highest weight is 
$n$ times that of the fundamental spinor.  
\eqn\agdbas{\eqalign{  
& Z_{\mu_1 \mu_2 \cdots \mu_s } 
= { 1 \over n^{s} } \sum_{r_1 \ne r_2 \cdots  \ne r_s } 
\rho_{r_1} ( \Gamma_{\mu_1} )  \rho_{r_2} ( \Gamma_{\mu_2} ) 
\cdots \rho_{r_s} ( \Gamma_{\mu_s} ) \cr 
} }
Each $r_i$ index can run from $1$ to $n$. 
The $\mu$ indices can take values from 
$1$ to $D+1$.    It is important to note 
that the maximum value that $s$ can take at finite $n$ is 
$n$.

The operators of the form \agdbas\ contain the symmetric 
traceless representation corresponding to Young Diagrams 
of type $ (s , 0, \cdots  )$ where the  integers denote the 
first, second etc.  row lengths. By contracting the indices, 
the operators of fixed $l$ also contain lower representations, 
e.g $(s-2, 0 ) $. For example 
\eqn\twocont{\eqalign{  
&  \sum_{ \mu}  Z_{ \mu \mu } = { n ( n-1)\over n^2}  \cr 
& \sum_{ \mu}  Z_{ \mu \mu  \alpha } = { ( n-1 )( n-2) \over n^2 } Z_
{\alpha} \cr   
 } } 
 We now define $ \cAn ( R^{D+1 }  )$ we keep the product
structure from $ \cAn ( S^D )$ but we drop relations 
between $Z$'s with contracted indices and $Z$'s with fewer 
indices such as \twocont. Note that $ \cAn ( R^
{D+1}  )$ is still finite
dimensional because $ s \le n $, but it a larger algebra that 
$ \cAn ( S^{D}  )$. We can obtain $ \cAn ( S^{D})$ from $ \cAn ( R^{D+1}  ) $  
 by imposing the contraction relations. 
The operators of the form given in \agdbas\ can be obtained 
by multiplying $ Z_{\mu_1} $ using the product in $ \cAn ( S^{D} )$. 
For example 
\eqn\genzmmz{ 
Z_{ \mu_1 } Z_{\mu_2 } = Z_{ \mu_1 \mu_2 } + { 1 \over n } { \delta_{\mu_1 \mu_2  } }}
More generally we have 
\eqn\zandpr{ 
Z_{\mu_1 \mu_2 \cdots  \mu_s } 
=  Z_{\mu_1} Z_{\mu_2} \cdots Z_{\mu_s } + { \cal{  O } } ( 1/n ) 
}
To make clear which product we are using we write 
\eqn\genmkcl{ 
m_2 ( Z_{ \mu_1 }  \otimes Z_{\mu_2 }  ) 
= Z_{ \mu_1 \mu_2 } + { 1 \over n } { \delta_{\mu_1 \mu_2 } }
} 
The $m_2$ denotes the product obtained by taking the Matrix product 
in $End ( \cRn )$ and projecting out the representations 
which do not transform in the symmetric traceless representations. 
This kind of product can be used for both $ \cAn ( R^{D+1} ) $ 
and $ \cAn ( S^D )$. Since the higher $Z$'s are being generated 
from the $ Z_{\alpha}$ we can view $ \cAn ( R^{D+1} )$ as a deformation of 
the algebra of polynomials in the $D+1$  variables $ Z_{\alpha}$. 
As discussed previously in the context of $ \cAn ( S^{D} )$
 \refs{ \sphdiv , \hdim }  this product is 
commutative and non-associative but becomes commutative and 
associative in the large $n$ limit.

Now we look at a deformed  $R^{2k+1} $  subspace 
of the deformed $R^{D+1}$ space. 
We have generators $ Z_{\mu_1 \mu_2 \cdots \mu_s}$ 
where the $ \mu $ indices take values from $1 $ to $2k+1$. 
 This operator is symmetric 
under any permutation of the $\mu$'s. 
The largest representation of $SO(2k+1)$ contained 
in the set of operators for fixed $s$ is 
the one associated with Young Diagram of row lengths 
$(r_1,r_2, \cdots ) = (s,0, \cdots )$ and   are symmetric traceless
representations.  We keep  $k$ fixed  and let $D$ be very large.  
We will get  very simple algebras which we denote as $ \cBn ( R^{2k+1} ) $ 
and $ \cBn ( S^{2k} ) $. The relation between  $ \cBn ( R^{2k+1} ) $
and  $ \cBn ( S^{2k} ) $ is similar to that between $ \cAn ( R^{2k+1} )$ 
and $ \cAn( S^{2k+1} )$.

Let us look at some simple examples of the product. 
\eqn\simpex{\eqalign{ 
X_{\mu_1} X_{\mu_2} &= \sum_{r_1=1}^{n}   \rho_{r_1} ( \Ga_{\mu_1} ) 
  \sum_{r_2=1}^{n} \rho_{r_2} ( \Ga_{\mu_2} ) \cr 
&= \sum_{r_1 \ne r_2 = 1 }^{n} \rho_{r_1} ( \Ga_{\mu_1}) \rho_{r_2}
( \Ga_{\mu_2} ) + \sum_{r_1=r_2=1}^{n} \rho_{r_1} 
 ( \Ga_{\mu_1}  \Gamma_{\mu_2} ) \cr 
& = \sum_{r_1 \ne r_2 = 1 }^{n} \rho_{r_1} ( \Ga_{\mu_1} ) \rho_{r_2}
( \Ga_{\mu_2} ) + \sum_{r_1=r_2=1}^{n} \rho_{r_1} 
 ( \delta_{\mu_1 \mu_2}  ) \cr 
& = X_{\mu_1 \mu_2} + n  \delta_{\mu_1 \mu_2} }} 
In the first line we have written the product of two 
$X$ operators which are both $\Gamma$ matrices 
acting on each of $n$ factors of a symmetrized tensor 
product of spinors \clt. 
In the second line we have separated the double sum into 
terms where $r_1=r_2$ and where they are different. 
In other words we are looking separately at terms where 
the two $\Gamma$'s act on the same tensor factor and when they act on
different tensor factors. Where $r_1 \ne r_2$ the expression 
is symmetric under exchange of $\mu_1 $ and $\mu_2$. When 
$r_1 =r_2$, there is a symmetric part and an antisymmetric 
part. The antisymmetric part is projected out when we 
want to define the product which closes on the 
fuzzy spherical harmonics \sphdiv. This is why the 
third line only keeps $\delta_{ \mu_1 \mu_2} $. 
Expressing the product in terms of the normalized $Z_{\mu}$ we have 
\eqn\normzp{ 
Z_{\mu_1} Z_{\mu_2} = Z_{\mu_1 \mu_2} + { 1 \over n } \delta_{\mu_1
\mu_2} } 
By similar means we compute 
\eqn\nxtprod{ 
 Z_{\mu_1 \mu_2}  Z_{ \mu_3} 
= Z_{\mu_1 \mu_2 \mu_3 } + { (n-1) \over n^2 } ( \delta_{\mu_1 \mu_3}
Z_{\mu_2} + \delta_{\mu_2 \mu_3} Z_{\mu_1} )  }
The LHS contains sums of the form 
\eqn\sepsums{ 
 \sum_{r_1 \ne r_2 } \sum_{r_3}
 =  \sum_{r_1 \ne r_2  \ne r_3 } +  \sum_{ r_3 = r_1 \ne r_2   } + 
\sum_{ r_1 \ne r_2 =r_3   } }
The three types of sums in \sepsums\ lead, respectively,
  to the first, second and third terms
on the RHS of \nxtprod. The factor of $n-1$ in the second term, for example, 
comes from the fact that the $r_1=r_2$ sum runs from $1$ to $n$ 
avoiding the single value $r_3$. The $1/n^2$ comes from
normalizations. The relations \normzp\ and \nxtprod\ hold both 
in $ \cBn ( R^{2k+1} ) $ and $ \cAn ( R^{2k+1} )$. 

Using \normzp\ and \nxtprod\ it is easy to see the non-associativity. 
Indeed 
\eqn\simpnass{\eqalign{ 
( Z_{\mu_1} Z_{\mu_2} ) Z_{\mu_3} &=
Z_{ \mu_1 \mu_2 \mu_3 } + { (n-1) \over n^2 } ( \delta_{\mu_1 \mu_3}
Z_{\mu_2} + \delta_{\mu_2 \mu_3} Z_{\mu_1} ) +  
{ 1 \over n } \delta_{\mu_1 \mu_2} Z_{\mu_3} \cr 
 Z_{\mu_1} ( Z_{\mu_2}  Z_{\mu_3} )
 & = Z_{ \mu_1 \mu_2 \mu_3 }  + { (n-1) \over n^2 } ( \delta_{\mu_1 \mu_3}
Z_{\mu_2} + \delta_{\mu_1 \mu_2 } Z_{\mu_3} ) + 
{ 1 \over n } \delta_{\mu_2 \mu_3} Z_{\mu_1} \cr 
( Z_{\mu_1} Z_{\mu_2} ) Z_{\mu_3} - Z_{\mu_1} ( Z_{\mu_2}  Z_{\mu_3}) 
&= { 1 \over n^2 }   ~~ ( ~~  \delta_{\mu_1 \mu_2} Z_{\mu_3} -  
\delta_{\mu_2 \mu_3} Z_{\mu_1}~~  )  \cr  }}

We now explain the simplification that arises 
in $ \cBn ( R^{2k+1} ) $ as opposed to $ \cAn ( R^{2k + 1} ) $.
Consider a product 
\eqn\prodtwtw{  
  Z_{ \mu_1 \mu_2 } Z_{\mu_3 \mu_4 } 
 = { 1 \over n^4 } \sum_{r_1 \ne r_2 } \rho_{r_1 } (  \Gamma_{\mu_1 } ) 
   \rho_{r_2} ( \Gamma_{\mu_2} ) .  \sum_{r_3 \ne r_4 } \rho_{r_3 } 
  (  \Gamma_{\mu_3 } ) \rho_{r_4} ( \Gamma_{\mu_4} ) }  
In $ \cAn ( R^{2k+1} ) $ or $ \cAn ( S^{2k} )$, 
we will get terms of the form $ Z_{\mu_1 \mu_2 \mu_3 \mu_4} $ 
where $r_1 \ne r_2 \ne r_3 \ne r_4 $. In addition there 
will be terms of the form $ \delta_{\mu_1 \mu_3 } Z_{ \mu_2 \mu_4 } $
from terms where we have $ r_1 = r_3 $ and $ r_2 \ne r_4 $. 
If $ r_1 = r_3 \ne r_2 \ne r_4 $, the antisymmetric part of 
$ \Gamma_{\mu_1} \Gamma_{\mu_3}$ appears in the Matrix product 
of $Z_{\mu_1 \mu_2} $ with $ Z_{\mu_3 \mu_4 } $ but transforms 
as the representation $ \vec r = ( 2, 1, 0, ...  ) $. Hence these 
are projected out. However when we consider terms 
coming from $ r_1 =r_3 $ and $r_2 = r_4 $ and take the 
operators of the form $ \rho_{r_1} ( [ \Gamma_{\mu_1} , \Gamma_{\mu_3 } ]  ) 
\rho_{r_2} (  [ \Gamma_{\mu_2} , \Gamma_{\mu_4} ]  ) $, these will 
include representations of type $ ( 2,2, 0, \cdots )$ 
which are projected out, in addition to those of type 
$(  2, 0 , 0 \cdots ) $. The latter come from 
the trace parts $ \delta_{\mu_1 \mu_2 }$. They have coefficients 
that go as $ {1 \over D} $ and hence disappear in the large $D$ limit. 
We conjecture that 
all terms of this sort, which transform as symmetric traceless
reps, but come from antisymmetric parts of products in 
coincident $r$ factors, are subleading in a $1/D $ expansion. 
The simple product on the  $ Z_{\mu_1 \mu_2 \cdots } $, 
 obtained by  dropping all the antisymmetric parts  
 in the coincident $r$ factors ( or according to the conjecture, 
by going to the large $D$ fixed $k$ limit )
gives an algebra we denote as $ \cBn ( R^{2k +1 } )$.  
We will not try to give a proof of the conjecture that 
$   \cBn ( R^{2k +1 } )$ arises indeed in the large $D$ limit 
$ \cAn ( R^{D+1} ) $ 
as outlined above. We may also just study $ \cBn ( R^{2k+1} ) $ 
as an algebra that has many similarities with 
$ \cAn ( R^{2k+1} ) $ but has simpler structure constants, while 
sharing the key features of being a commutative non-associative 
algebra which approaches the ordinary polynomial algebra 
of functions on $R^{2k+1}$ in the large $n$ limit.

We now give an explicit description of the product 
on elements  $Z_{\mu_1 \cdots \mu_s}$ in the algebra $ \cBn ( R^{2k+1})$ .  
It will be convenient to set up some definitions 
as a prelude to a general formula. For a set of integers 
$S$ we will denote by $ Z_{\mu ( S ) } $ an operator
of the form \agdbas\ with the labels on the $\mu$'s taking values 
in the set $S$. For example $Z_{\mu_1 \cdots \mu_s}$ is of the 
form $ Z_{\mu ( S ) } $ 
with the set $S$ being the set of integers ranging from 
$1 $ to $s$. Instead of writing $ Z_{\mu_1 \mu_4 \mu_5 } $ 
we can write $ Z_{ \mu( S ) } $ with $S$ being the set 
of integers $ \{ 1, 4, 5 \} $.  

Given two sets of  positive integers $T_1$ and $T_2$ 
where $T_1$ and $T_2$ have the same number of elements, 
which we denote as $ |T_1| = |T_2| \equiv t $, we define 
\eqn\defdelt{ 
\tilde \delta ~~ ( ~~ \mu ( T_1 ) ~ \mu (T_2 ) ~~ ) = \sum_{ \sigma }  
\delta ( ~\mu_{i_1} ~ \mu_{  j_{ \sigma(1) } }~ )
\cdots \delta ( ~ \mu_{i_{ t }} ~~ \mu_{  j_{ \sigma ( t  ) }  } ~   )
}    
The $\delta$'s on the RHS are ordinary Kronecker deltas. 
The $i$'s are the integers of the set $T_1$ with a fixed ordering, 
which can be taken as $ i_1  < i_2 < \cdots < i_{ t }  $. 
The $j$'s are the integers of the set $T_2$ with a similar ordering. 
The $\sigma$ in the sum runs over all the permutations in the group 
of permutations of $ t$ elements. The $\tilde \delta $ 
is therefore  a sum over all the $ t ! $ ways of pairing 
the integers in the set $T_1$ with those in the set  $T_2$. 

With this notation the general formula for the product 
can be expressed as
\eqn\strcprod{\eqalign{ 
& Z_{\mu ( S_1 ) } Z_{\mu ( S_2 ) } 
= \sum_{ |T_1| } \sum_{ |T_2 | } ~~~ \delta ( ~|T_1| , |T_2| ~ ) ~~~ 
  { \sum_{T_1 \subset S_1 } } \sum_{T_2 \subset S_2 }    \cr 
&  {1 \over n^{2 |T_1| } }  ~~ { ( n - |S_1| - | S_2| + 2 |T_1| ) ! \over    
( n - |S_1| - | S_2| +  |T_1| ) !  } 
~~ \tilde \delta (~  \mu ( T_1 ) ~,~ \mu ( T_2 ) ~ ) 
   ~~~ Z_{ \mu (   S_1 \cup S_2  \setminus { T_1 \cup  T_2  } ) }
\cr 
}}   
We have chosen $S$ and $T$ to be the sets 
describing two elements of $ \cBn$ of the form 
\agdbas.  
There is a sum over positive integers $|T_1|$
and $|T_2|$ which are the cardinalities of 
subsets $T_1$ and $T_2$ of $S $ and $T$ respectively.  
Given the restriction 
$ |T_1| = |T_2|$, the sum over $ |T_1| $ extends 
from $ 0$ to  $ min ( |S| , |T| )$.  
The factor expressed in terms of factorials 
comes from the $ |T_1|$ summations which can be done after 
replacing quadratic expressions in $ \Gamma$'s by 
 $\delta$'s. The formula expresses the fact that the 
different terms in the product are obtained by summing 
over different ways of picking subsets $T_1$ and $T_2$ 
of $S$ and $T$ which contain the elements that lead to 
$\delta$'s. The set of remaining elements 
$ S \cup T  \setminus { T_1 \cup  T_2  }$  gives 
an operator of the form \agdbas. It is instructive to check that 
\normzp\ and \nxtprod\ are special cases of \strcprod. 
Note that  we should set 
to zero all  $ Z_{\mu ( S) } $ where $ |S| > n $. 
For example this leads to a restriction on the sum over  $ |T_1| $ 
requiring it to start at  $ max ( 0, |S| + |T| - n )$. 
This is not an issue if we are doing the $1/n$ expansion.

\newsec{ Deformed derivations  on $\cBn ( R^{2k+1}  )$ }

We now define operators $ \delta_{\alpha}$,  which
are derivations in the classical limit, 
by giving their action on the above basis. 
\eqn\acbas{
\delta_{\alpha}  Z_{\mu_1 \mu_2 \cdots \mu_s } 
= \sum_{i=1}^{s} \delta_{\alpha \mu_i} Z_{\mu ( 1..s \setminus i )}  
} 

At finite $n$ these are not 
derivations. Rather they are deformed 
derivations, which can described in terms 
of a co-product, a structure which  comes up for example 
in describing the action of quantum enveloping algebras on tensor products 
of representations or on products of elements of a $q$-deformed 
space of functions ( see for example \refs{ \maj, \charpress }  )  
\eqn\defcop{\eqalign{  
\Delta( \delta_{\alpha}) &=  ( \delta_{\alpha} \otimes 1 + 1 \otimes
\delta_{\alpha} )  \sum_{k=0}^{\infty} { (-1)^k \over n^{2k}} ~~~~ 
\delta_{\alpha_1}
\delta_{\alpha_2} \cdots \delta_{\alpha_k} \otimes \delta_{\alpha_1}
\delta_{\alpha_2} \cdots \delta_{\alpha_k}   \cr 
  &= 
\sum_{k=0}^{\infty} { (-1)^k \over n^{2k}} ~~~~ 
\delta_{\alpha}\delta_{\alpha_1}
\delta_{\alpha_2} \cdots \delta_{\alpha_k} \otimes \delta_{\alpha_1}
\delta_{\alpha_2} \cdots \delta_{\alpha_k} \cr 
 &~~~~~~~~+ \sum_{k=0}^{\infty} { (-1)^k \over n^{2k}} ~~~~ 
\delta_{\alpha_1}
\delta_{\alpha_2} \cdots \delta_{\alpha_k} \otimes \delta_{\alpha} 
\delta_{\alpha_1}
\delta_{\alpha_2} \cdots \delta_{\alpha_k}
\cr  } }
The leading $k=0$ terms just lead to the 
ordinary Leibniz rule, and the corrections are subleading 
as $n \rightarrow \infty $. These formulae 
show that  co-product  is a useful 
structure for describing  the deformation of Leibniz 
rule. Another possibility one might consider is 
to see if adding $1/n$ corrections of the type 
 $ \delta_{\alpha} ( 1 + { 1 \over n^2}  \delta_{\beta } \delta_{\beta } ) $ 
gives a derivative which obeys the exact Leibniz rule. This latter 
possibility does not seem to work. 

The co-product  \defcop\ has the property 
that 
\eqn\cop{ 
\delta_{\alpha}  . m_2 = m_2 . \Delta ( \delta_{\alpha} ) 
} 
Another way of expressing this is that 
\eqn\copi{\eqalign{  
& \delta_{\alpha}  ( A . B )  = m_2 . \Delta ( \delta_{\alpha} ) A
\otimes  B 
\cr 
& =   \sum_{k=0}^{\infty} { (-1)^k \over n^{2k}} ~~~~
\delta_{\alpha} \delta_{\alpha_1}
\delta_{\alpha_2} \cdots \delta_{\alpha_k}  A  ~.~ 
\delta_{\alpha_1}\delta_{\alpha_2} \cdots \delta_{\alpha_k} B  \cr 
& + \sum_{k=0}^{\infty} { (-1)^k \over n^{2k}} ~~~~
 \delta_{\alpha_1}
\delta_{\alpha_2} \cdots \delta_{\alpha_k}  A ~.~ 
\delta_{\alpha} \delta_{\alpha_1}
\delta_{\alpha_2} \cdots \delta_{\alpha_k} B 
}}

The proof is obtained by evaluating 
LHS and RHS on arbitrary pair of elements 
in the algebra.

It is useful to summarize some properties of 
$\delta_{\alpha } $ which act as  deformed derivations 
on the  { \it deformed polynomial algebra } $ \cAn ( R^{2k+1}  )$  
They  approach $ \partial_{\alpha}$ in the large $n$ 
limit. 
At finite $n$ they  obey $ [ \delta_{\alpha} , \delta_{\beta} ] = 0$
At finite $n$ they  transform as a vector under the 
$SO(2k+1)$ Lie algebra of derivations that acts 
on $\cBn$, i.e we have
\eqn\spropder{ 
 [ L_{\mu \nu } , \delta_{\alpha} ] = \delta_{\nu \alpha}
\delta_{\mu} - \delta_{\mu \alpha} \delta_{\nu} } 
These properties can be checked by acting 
with the LHS ad RHS on a general element of the algebra.

\subsec{ A useful  identity  on binomial coefficients } 

In proving that the above is the right co-product
we find the following to be a useful identity. 

\eqn\numident{
 \sum_{k=0}^{A} (-1)^k 
{ (N+1 +2A )! \over (A-k)! ( N+1+A+k)! }= { (N + 2A  )!\over A! ( N+A)! }   } 

This is  a special case  of the following equation 
from \grrefi\ 
\eqn\frgr{ 
\sum_{r=0}^{n} (-1)^r { n \choose r }  { (r+b-1)! \over (r+a-1)! } 
= { ( n+a-b-1)! ( b-1)! \over (n+a-1)! (a-b-1)!} }
Choose $n = A $, $b=1$ and $ a = N+A+2 $ in \frgr\ 
to get precisely the desired equation \numident.

 The proof of the deformed derivation property 
 \copi\ proceeds by writing out both sides 
 of the equation for a general pair 
of elements $A$, $B$ of the form $ Z_{ \mu ( S ) } $ and 
$ Z_{ \mu ( T ) } $. 

For the LHS where we multiply first and take derivatives 
after, that is we compute $  \delta_{\alpha}.  m_2 . A \otimes B $
we have 
\eqn\expfdm{\eqalign{  
& \delta_{\alpha} ~ . ~  m_2 ~.~   Z_{ \mu ( S ) }  \otimes  Z_{ \mu (T) } 
 = \sum_{ |S_1| } \sum_{ |S_2| }  
 ~~~~ \sum_{S_1 \subset S  }   \quad \sum_{S_2 \subset T  }
\quad \sum_{i \subset \{ S \cup T \} \setminus \{ S_1 \cup S_2 \}  } 
  \cr 
&\delta ~(~  |S_1| ~,~ |S_2| ~)  \tlddel{\mu}{S_1}{S_2} 
\delta_{\alpha \mu_i }
 { ( (n- |S| - |T|  + 2 |S_1| )! \over ( n - |S| - |T| + |S_1| ) ! } 
Z_{ \mu ( \{ S \cup T \} \setminus \{ S_1 \cup S_2 \cup i \}  ) } 
} }

For the RHS of \copi\ we have  
\eqn\deltonzszt{\eqalign{  
&   m_2 ~.~ \Delta ( \delta_{ \alpha} ) ~.~  ( Z_{ \mu ( S ) } \otimes Z_
  { \mu ( T )} )  \cr 
& = \sum_{ |T_1| }  \sum_{ |T_2| }  \sum_{ |T_3| }  \sum_{ |T_4| } 
 ~~~ \delta ~(~~ |T_1| ~,~ |T_2| ~~ )  ~~~ \delta ~(~~ |T_3| ~,~ |T_4|
~~ ) \cr 
&  \sum_{ T_1 \subset  S ~~,~~ T_2 \subset T  }  \qquad
 \sum_{ T_3 ~ \subset ~ S ~\setminus~ T_1  ~~,~~ T_4 ~\subset ~T ~\setminus~ T_2  }
\qquad \sum_{i ~~\subset ~~ \{ ~ S \cup T ~\} ~\setminus ~\{  T_1 \cup
T_2 \cup T_3 \cup T_4 ~\}  }   
  \cr 
& (-1)^{ |T_1|} ~~ |T_1|! ~~ { ( n - |S| - |T| + 2|T_1| + 2 |T_3| +1 )! \over 
 ( n - |S| - |T| + 2 |T_1| + |T_3| + 1 )! } \cr 
& \delta_{ \alpha \mu_i } ~~ \tlddel{\mu}{T_1}{T_2} ~~ \tlddel{\mu}{T_3}{T_4}
~~ Z_{\mu( \{ S \cup T \} \setminus \{ T_1 \cup T_2 \cup T_3 \cup T_4
\cup i \}  ) }
\cr }}
This follows from the the definition of $ \Delta( \delta_{\alpha})$ 
given in \defcop, the action of $\delta_{\alpha}$ described 
in \acbas , and the general structure of the product described 
in \strcprod. The $\mu$ indices associated to the 
  pair of subsets $ (T_1, T_2)$ are contracted with Kronecker deltas 
coming from the tensor product $ \delta_{\alpha_1} \delta_{\alpha_2} 
\cdots \delta_{\alpha_{ |T_1| } } \otimes
  \delta_{\alpha_1} \delta_{\alpha_2} 
\cdots \delta_{\alpha_{ |T_1| } } $ appearing in $ \Delta
( \delta_{\alpha }) $. The $|T_1|! $ arises because  
the same set of $\mu$ contractions can come from different 
ways of applying the $\delta_{\alpha}$'s in  $ \Delta
( \delta_{\alpha} )$. The contractions involving the sets $T_3$ and $T_4$ 
of indices come from the multiplication $m_2$.  We will manipulate 
\deltonzszt\ to show equality with  \expfdm.

We write $ S_{1} = T_1 \cup T_3  \subset S $ and $S_2 = T_2 \cup T_4
\subset T $. We observe that the $Z$'s  in \deltonzszt\ 
only depend on these subsets. 
We have of course the relations $ |S_1| = |T_1| + |T_3| $ 
and $ |S_2| = |T_2| + |T_4| $ and $ |S_1| = |S_2|$.   For fixed subsets 
$S_1$, $S_2$ we observe the identity : 
\eqn\iddelstwo{\eqalign{ 
& \sum_{T_1 \subset S_1 ~,~ T_2 \subset S_2 } \qquad 
\sum_{T_3 \subset S_1\setminus T_1 ~,~ T_4 \subset S_2 \setminus T_2 }
\tlddel{\mu}{T_1}{T_2} \tlddel{\mu}{T_3}{T_4} \cr 
& = { ( |T_1| + |T_3| ) ! \over |T_1|! |T_3|! } \tlddel{\mu}{S_1}{S_2}
\cr }}
For fixed sets $S_1, S_2$, the $\tlddel{\mu}{S_1}{S_2}$ 
is a sum of $ |S_1|! $ terms. The sums on the LHS 
contain $ { |S_1| \choose |T_1| } { |S_2| \choose |T_2| }
= { |S_1| \choose |T_1| }^2 $ choices of $ T_1,T_2$, 
and the two $ \tilde \delta$'s on the LHS 
contain   a sum of $| T_1|! |T_2|! = (|T_1|!)^2$ 
terms, each being a product of Kronecker deltas. 
The combinatoric factor  in \iddelstwo\ is the ratio 
$ { |S_1| \choose |T_1| } { |S_2| \choose |T_2| } |T_1|! |T_3|! \over 
 |S_1|!$.
 Given that the summands can be written in terms of 
$S_1$ and $S_2$, 
the summations over the cardinalities of the  $T_i$ subsets
can be rewritten using 
\eqn\cardsconv{ 
\sum_{ |T_1| } \sum_{ |T_2| }\sum_{ |T_3| }\sum_{ |T_4| } 
~~~ \delta ( |T_1|, |T_2| ) ~~~ \delta ( |T_3|, |T_4| )  
= \sum_{ |S_1| } \sum_{ |S_2| } ~~~ \delta ( |S_1| , |S_2| ) 
} 

Using \iddelstwo\ and \cardsconv\ we can rewrite \deltonzszt\ 
as 
\eqn\rewritdoz{\eqalign{ 
& \sum_{|S_1|} \sum_{|S_2|} \delta ( |S_1|, |S_2| ) \sum_{ |T_1| = 0 }^
{ |S_1| }  ( -1 )^{ |T_1| } |T_1|!  {( |T_1| + |T_3| ) ! \over  
|T_1|! |T_3|! } \cr 
& \sum_{S_1 \subset S } \sum_{S_2 \subset T} \tlddel{\mu}{S_1}{S_2} 
\sum_{i\subset  \{ S \cup T \}  \setminus \{ S_1 \cup S_2 \}  }
 \delta_{\alpha \mu_i
} \cr 
& { ( N + 2|T_1| + 2 |T_3| +1 ) ! \over  ( N + 2|T_1| + |T_3| +1 )! }
Z_{\mu ( \{ S \cup T \}  \setminus \{ S_1 \cup S_2 \cup i \} ) } \cr }}
To simplify the combinatoric factors we have defined 
$N \equiv n - |S| - |T|$. 
We can further simplify as follows 
\eqn\frsm{\eqalign{ 
& \sum_{ |S_1| } ~~~ \bigl ( ~~~ \sum_{ |T_1| = 0 }^{ |S_1| } (-1)^{ |T_1| } 
 { |S_1|! \over (|S_1| - |T_1| ) !} { ( N + 2|S_1| +1 )! \over ( N +
|S_1|+ |T_1| +1 )!} ~~~ \bigr ) \cr  
&\sum_{S_1 \subset S}~~~~  \sum_{S_2 \subset T } ~~~~~~ 
\sum_{ i \subset { \{ S
\cup T \} }
\setminus \{ S_1 \cup S_2 \}  } \tlddel{\mu}{S_1}{S_2} \delta_{\alpha \mu_i} 
Z_{ \mu  ( \{ S \cup T \} \setminus \{ S_1 \cup S_2 \cup i \}  ) } \cr 
&   = \sum_{ |S_1| } { ( N + 2 |S_1| ) ! \over ( N + |S_1| ) ! } 
\sum_{S_1 \subset S} ~~~~ \sum_{S_2 \subset T } ~~~~~~ \sum_{ i \subset { \{ S
\cup T \} }
\setminus \{ S_1 \cup S_2 \}  } \cr
& \tlddel{\mu}{S_1}{S_2} \delta_{\alpha \mu_i} 
Z_{ \mu  ( \{ S \cup T \} \setminus \{ S_1 \cup S_2 \cup i \}  ) } \cr 
}}
In the final equality we have used 
\numident\ by setting $A = |S_1| $ and $ k = |T_1|$.  
This establishes the equality of \expfdm\ and \deltonzszt.

\newsec{ Derivations for the twisted algebra $\cBnst ( R^{2k+1})$   }

We defined above certain deformed derivations.  
This is reminiscent of quantum groups where 
the quantum group generators act on tensor 
products via a deformation of the usual 
co-product. In this context, classical and quantum co-products 
can be related  by a Drinfeld twist \drinf, in which  
an element $F$ living in $ U \otimes U$ plays a role,  
where $U$ is the quantum enveloping algebra. 
This suggests that we could twist the 
product $m_2$ of the algebra $\cBn ( R^{2k+1})$ 
to get a new product $m_2^*$ which defines a new algebra 
 $\cBnst ( R^{2k+1})$. These two algebras share the same underlying vector 
space. 
We want to define a star product $m_2^*$ 
which is a twisting of the product $m_2$, 
for which the $ \delta_{\alpha}$ 
are really  derivations rather than deformed 
derivations.  It turns out that this becomes possible after 
we use in addition to the $ \delta_{\alpha}$ a degree operator $D$. 
For symmetric elements  of the form $ Z_{ \mu  ( S ) } $ 
the degree operator is defined to have eigenvalue $ |S | $. 
In the following we will describe a twist which uses 
$D, \delta_{\alpha}$ and leads to a product $m_2^*$ 
for which the $  \delta_{\alpha}$ are really derivations. 
It will be interesting to see if the formulae given here 
are related to Drinfeld twists in a precise sense. 
The physical idea is that field theory actions can be written equivalently either 
using $m_2$ or $m_2^*$, somewhat along the lines of the Seiberg-Witten 
map \seibwit.

\subsec{ Formula for The star product } 

Let us write $m_2^* $ for the star product. It is useful to 
view $m_2^*$ as a map from $ \cAn \otimes \cAn $ to $\cAn $. 
With respect to the new product $m_2^*$, 
$\delta_{\alpha}$  obey the Leibniz rule at finite $n$. 
\eqn\leibfrst{ 
\delta_{\alpha} ~~ . ~~  m_2^* = m_2^* ~  . ~   ( ~  \delta_{\alpha} \otimes
1 ~ + ~  1 \otimes \delta_{\alpha} ~ )  }
Equivalently for two arbitrary elements $A$ and $B$ of the algebra, we 
have, after writing $ m_2^* ( A \otimes B ) = A * B $,   
\eqn\leibst{ 
 \delta_{\alpha} ( A * B ) =  ( \delta_{\alpha} A )  *  B ~ + ~  
A *  ( \delta_{ \alpha } B )  }

We look for a formula for 
$m_2^*$ as an expansion in terms of 
$m_2$,  the derivatives $ \delta_{\alpha}$, 
and using a function of the degree operator to be determined.  
The ansatz takes the form 
\eqn\anstz{ 
m_2^* =  \sum_{l=0}^{\infty}   ~~ { 1\over n^{2l}}~~
  h_l ( D ) ~   .  ~  m_2  ~ .  ~ 
\delta_{\alpha_1}\delta_{\alpha_2}  \cdots \delta_{\alpha_l} 
\otimes \delta_{\alpha_1}\delta_{\alpha_2}  \cdots \delta_{\alpha_l} 
}
The function $h_l(D)$ is determined by requiring 
that \leibst\ hold for arbitrary $A$ and $B$. 
It turns out, as we show later, that 
\eqn\detf{ 
 h_l ( D ) =  { D \choose l } } 
For finite $n$ the degrees of the operators 
do not exceed $n$. We can hence restrict the 
range of summation in \anstz\ from $0$ to $n$. 

A useful identity in proving that with the choice 
\detf\ the product in \anstz\ satisfies the Leibniz rule
is 
\eqn\combfrst{ 
\sum_{l=0}^{p_1}   h_{l} ( s-1 )  { n-s+1 \choose p_1 - l } 
= \sum_{l=0}^{p_1} h_{l} ( s )  { n-s \choose p_1 - l }
} 
This follows from  the combinatoric identity  ( see for example
\grrefii\ ) which can be written in the suggestive form 
\eqn\combid{ 
\sum_{l=0}^{p_1 }  { N \choose k } { M \choose p_1 -k } = { N + M \choose p_1
}} 
Substituting $ N $ with $ s-1$ and $M$ with $n-s+1$ we have 
\eqn\frlft{ 
 \sum_{l=0}^{p_1}  {  s-1  \choose l }   { n-s +1 \choose p_1 - l }
 = { n \choose p_1 } 
} 

Substituting $ N$ with $s $ and $M$ with $n-s$ 
we have 
\eqn\frlft{ 
 \sum_{l=0}^{p_1}  {  s  \choose l }   { n-s \choose p_1 - l }
 = { n \choose p_1 } 
} 
Hence the identity \combfrst\ required of $f$ is indeed 
satisfied by the choice \detf.

\subsec{ Proof of the derivation property for the twisted product } 

We compute $ \delta_{\alpha} . m_2^* . ( Z_{\mu(S)} \otimes Z_
{ \mu ( T) } ) $ using the definitions \anstz , \acbas\ and the 
general form of the $m_2$ product in \strcprod,to obtain 
\eqn\dmtst{\eqalign{ 
 & \sum_{ l =0 }^{ \infty } \sum_{ |T_1| }  \sum_{ |T_2| } \delta
( |T_1| , l  )  \delta ( |T_2| , l  )  \sum_{|T_3|  } \sum_{T_4| } 
\delta ( |T_3| , |T_4| )  \cr 
& \sum_{T_1 \subset S } ~~~ \sum_{ T_2 \subset T }  ~~~~ \sum_{ T_3 \subset S
 \setminus T_1 } ~~~~  \sum_{ T_4 \subset T \setminus T_2  } 
\sum_{ i \in  \{ S \cup T \} \setminus \{ T_1 \cup T_2 \cup T_3 \cup T_4 \} }  
 \cr 
&  { l ! \over n^ { 2l } } h_{l} ( |S| + |T| - 2 l  - 2 |T_3| ) 
 { 1 \over n^{ 2 |T_3| } } 
   { ( n - |S| - |T|  + 2 l  + 2 |T_3| ) ! \over  ( n - |S| - |T|
 + 2  l  +  |T_3| ) ! }  \cr 
& \tlddel{\mu}{T_1}{T_2} ~~~ \tlddel{\mu}{T_3}{T_4} ~~ \delta_{\mu
 \alpha_i } ~~ Z_{ \mu ( \{ S \cup T \} \setminus \{ T_1 \cup T_2 \cup T_3
 \cup T_4 \cup i \} )  } \cr }}

Similarly we  compute 
$ m_2^* . ( \delta_{\alpha} \otimes  1 + 1 \otimes \delta_{\alpha} )
. ( Z_{\mu(S) } \otimes Z_{ \mu (T) } ) $ to obtain 
\eqn\mstd{\eqalign{  
& \sum_{l=0}^{\infty} 
  \sum_{ |T_1| }  \sum_{ |T_2| } \delta
( |T_1| , l  )  \delta ( |T_2| , l  )  \sum_{|T_3|  } \sum_{T_4| } 
\delta ( |T_3| , |T_4| ) \cr 
& \bigl ( \sum_{j\in S } \sum_{T_1 \subset S\setminus j } ~~ \sum_{T_2 \subset T } ~~~ 
 \sum_{ T_3 \subset S \setminus j \cup T_1 } ~~~ 
\sum_{ T_4 \subset T \setminus T_2 } \cr 
& + \sum_{j\in T } \sum_{T_1 \subset S } ~~ \sum_{T_2 \subset T
\setminus j  } ~~~  
 \sum_{ T_3 \subset S \setminus  T_1 } ~~~ 
\sum_{ T_4 \subset T \setminus j \cup T_2 }  \bigr ) \cr 
& \delta_{ \alpha \mu_j } ~~\tlddel{\mu}{T_1}{T_2} ~~~ \tlddel{\mu}{T_3}{T_4}
Z_{\mu ( \{ S \cup T \}  \setminus \{ j \cup T_1 \cup T_2 \cup T_3
\cup T_4 \} )} \cr 
& { l! \over n^{2l} } { 1 \over n^{2 |T_3| }  }
 { ( n - |S| - |T|  + 2 l  + 2 |T_3| +1  ) ! \over ( n - |S| - |T| +
2l + |T_3| +1 ) ! } h_{l} ( |S| + |T| - 2 |T_1| - 2 |T_3| -1 ) 
}} 

At finite $n$ the number of linearly independent operators 
$ Z$ we need to consider are bounded by $n$, i.e $ |S| \le n $ and 
$ |T| \le n$. This means that the sum over $l$ is also a finite sum. 

Observe in \mstd\ that the $Z$ only depends on the 
unions $ T_1 \cup T_3 $ and $T_2 \cup T_4 $. We denote 
$ S_1 \equiv T_1 \cup T_3 $ and $S_2 = T_2 \cup T_4 $. 
It follows that the cardinalities are related 
$|S_1| = |T_1| + |T_3|$ and $ |S_2 | = |T_2| + |T_4|$.   
The sums over subsets in \mstd\ can be rearranged as 
\eqn\earmstd{ 
\sum_{ |S_1| } \sum_{ |S_2| } \delta ( |S_1| , |S_2 | ) 
\sum_{ S_1 \subset S } \sum_{ S_2 \subset T } \sum_{ j \in 
\{ S \cup T \} \setminus \{ S_1 \cup S_2  \} }}
Hence we re-express \mstd\ as 
\eqn\nmstd{\eqalign{ 
&  \sum_{ |S_1| } \sum_{ |S_2| } \delta ( |S_1| , |S_2 | ) 
\sum_{ S_1 \subset S } \sum_{ S_2 \subset T } \sum_{ j \in 
\{ S \cup T \} \setminus \{ S_1 \cup S_2 \}}  \sum_{ l=0 }^{ |S_1| }
\cr 
& \tlddel{\mu}{S_1}{S_2} ~~ \delta_{\alpha \mu_{j}} Z_{ \mu( \{ S \cup T \} 
\setminus \{ S_1 \cup S_2 \cup j \}  )}  \cr 
& { ( n - |S| - |T| + 2 l + 2 |T_3| + 1 ) ! \over 
( n - |S| - |T| + 2 l +  |T_3| + 1 ) ! }~~ l! ~~~  
{1 \over n^{2 |T_3| + 2 l  }}  ~~~~
{ |S_1|! \over l! ( |S_1| - l ) ! } \cr 
}}
The binomial factor ${ |S_1|! \over l! ( |S_1| - l ) ! }$
appears in using the conversion \iddelstwo\ of a sum of  products 
of $\tilde \delta$'s to a single $\tilde \delta $. 
Simplifying \earmstd\ 
\eqn\searmstd{\eqalign{  
&  \sum_{ |S_1| } \sum_{ |S_2| } \delta ( |S_1| , |S_2 | ) 
\sum_{ S_1 \subset S } \sum_{ S_2 \subset T } \sum_{ j \in 
\{ S \cup T \}  \setminus \{ S_1 \cup S_2  \} } 
\cr 
& { 1 \over n^{ 2 |S_1|} } \tlddel{\mu}{S_1}{S_2} 
\delta_{\alpha \mu_j} Z_{ \mu( \{ S \cup T \} 
\setminus \{ S_1 \cup S_2 \cup j \} )} \cr 
& \bigl ( \sum_{ l=0 }^{|S|_1 } { ( n - |S| - |T| + 2 |S_1| + 1 ) ! \over 
( n - |S| - |T| + |S_1|  +  l + 1 ) ! } h_{l} ( |S| + |T| - 2|S_1| -1
) { |S_1|! \over (|S_1| - l )! }  \bigr ) \cr 
}}

Similar manipulations can be done starting from \dmtst\ 
to collect the products of $\tilde \delta$'s into 
a single $ \tilde \delta $. We end up with 
\eqn\endup{ \eqalign{ 
&  \sum_{ |S_1| } \sum_{ |S_2| } \delta ( |S_1| , |S_2 | ) 
\sum_{ S_1 \subset S } \sum_{ S_2 \subset T } \sum_{ j \in 
\{ S \cup T \} \setminus \{ S_1 \cup S_2 \}   } 
\cr 
& { 1 \over n^{ 2 |S_1|} } \tlddel{\mu}{S_1}{S_2} 
\delta_{\alpha \mu_j} Z_{ \mu( \{ S \cup T \} 
\setminus \{ S_1 \cup S_2 \cup j \} )} \cr 
&\bigl ( \sum_{ l=0 }^{|S|_1 } { ( n - |S| - |T| + 2 |S_1|  ) ! \over 
( n - |S| - |T| + |S_1|  +  l  ) ! } h_{l} ( |S| + |T| - 2|S_1| 
) { |S_1|! \over (|S_1| - l )! }  \bigr ) \cr 
}}
This expression is the same as \searmstd\ except 
in the sum appearing in the last line. Since \searmstd\ 
has the derivative acting before the $m_2^*$,  the $h_l ( D )$ 
in the $ m_2^* $ has an argument which is the degree 
$ |S| + |T| - 2 |S_1| -1 $. This is because $ Z_{ \mu ( S \cup T ) } $
has degree $ |S| + |T| $ and the degree gets reduced by $ 2 |S_1| = 2
|T_1| + 2 |T_3| $ 
due to the $|T_1|$ contractions from the derivatives in $m_2^*$
and the $|T_3|$ contractions from the product $m_2 $ in the expression 
for $m_2^*$. Finally the reduction by $1$ is due to the
$\delta_{\alpha}$ which has already acted before the $m_2^*$ 
and hence before $ h_l ( D ) $. In \endup\ 
the argument of $h_l$ is larger by $1$ because  $\delta_{\alpha}$ 
acts before $m_2^*$. The  ratio of factorials 
also has a relative shift of $1$ because they arise from 
$m_2$ according to \strcprod\ and the $m_2$ acts before 
the $\delta_{\alpha}$ in one case and after it in the other. 
The validity  of \leibfrst\ or equivalently the equality 
of \dmtst\ and \mstd\ will follow from the equality of the sums 
\eqn\eqlsms{\eqalign{  
& \sum_{ l=0 }^{|S_1| } { ( n - |S| - |T| + 2 |S_1|  ) ! \over 
( n - |S| - |T| + |S_1|  +  l  ) ! } h_{l} ( |S| + |T| - 2|S_1| 
) { |S_1|! \over (|S_1| - l )! } \cr 
& = \sum_{ l=0 }^{|S|_1 } { ( n - |S| - |T| + 2 |S_1| + 1 ) ! \over 
( n - |S| - |T| + |S_1|  +  l + 1 ) ! } h_{l} ( |S| + |T| - 2|S_1| -1
) { |S_1|! \over (|S_1| - l )! } \cr 
}} 
Substituting $ h_l ( D ) = { D \choose l } $ 
and defining $u = |S| + |T| - 2 |S_1| $ to simplify formulae, 
the LHS of \eqlsms\ becomes  
\eqn\eqlsmsi{\eqalign{  
& \sum_{ l=0 }^{|S|_1 } { ( n - u  ) ! \over 
( n - u - |S_1|  +  l  ) ! }     { u! \over l! (u-l)! } 
 { |S_1|! \over (|S_1| - l )! } \cr 
& = |S_1|! \sum_{ l=0 }^{|S_1| } { n - u \choose |S_1| - l } { u \choose
l } \cr 
& = { n! \over  ( n - |S_1|) ! }  \cr 
}}
In the last step we used \combid\ ( see for example \grrefii\ ) 
with $N = u $ and $M = (n-u) $. 
For the RHS of \eqlsms\ we get 
\eqn\eqlsmsii{\eqalign{  
& \sum_{ l=0 }^{|S_1| } { ( n - u +1  ) ! \over 
( n - u - |S_1|  +  l +1   ) ! }     { (u-1) ! \over l! (u-l-1)! } 
 { |S_1|! \over (|S_1| - l )! } \cr 
& = |S_1|! \sum_{ l=0 }^{|S_1| } { n - u +1 \choose |S_1| - l } { u -1
\choose
l } \cr 
& = { n! \over  ( n - |S_1|) ! } \cr 
}}
Here we have used \combid\ with $ N = u -1 $ and $ M = n- u + 1 $. 
This establishes the equation \leibfrst, and 
also shows that \dmtst\ and \mstd\ can be simplified to 
\eqn\simpmstd{\eqalign{
&  \sum_{ |S_1| } \sum_{ |S_2| } \delta ( |S_1| , |S_2 | ) 
\sum_{ S_1 \subset S } \sum_{ S_2 \subset T } \sum_{ j \in 
\{ S \cup T \}  \setminus \{ S_1 \cup S_2  \} } 
\cr 
& { 1 \over n^{ 2 |S_1|} } \tlddel{\mu}{S_1}{S_2} 
\delta_{\alpha \mu_j} Z_{ \mu( \{ S \cup T \} 
\setminus \{ S_1 \cup S_2 \cup j \}  )} 
{ n ! \over ( n - |S_1| ) ! } \cr 
}}

It is also useful to note, as a corollary, that 
\eqn\fzmusmut{\eqalign{  
& Z_{ \mu ( S) } * Z_{ \mu ( T) } =
\sum_{ | S_1|  } \sum_{ | S_2|  } \delta ( |S_1| , |S_2 | )
  \sum_{ S_1 \subset S } \sum_{ S_2 \subset T }  
 { n! \over ( n- |S_1| ) ! } { 1 \over n^{2|S_1|} } \cr 
& \qquad \qquad \qquad  \qquad  
\tlddel{\mu}{S_1}{S_2}  Z_{\mu ( S \cup T \setminus S_1
\cup S_2 ) } \cr 
}}

\newsec{ Towards abelian gauge theory for $ \cBnst ( R^{2k+1} ) $   }

\subsec{ Gauge transformations and non-associativity} 

The construction of gauge theory starts with the 
definition of covariant derivatives
$ D_{\alpha} = \delta_{\alpha} - i A_{\alpha}$. 
These are used to define covariant field strengths 
which lead to invariant actions. 
Consider some field
$\Phi$ in the fundamental of the gauge group. Under a gauge
transformation by a gauge parameter $\epsilon $ the variation is 
given, as usual, by 
\eqn\basic{           
  \delta \Phi = i \epsilon  * \Phi } 
The desired covariance property of the proposed  covariant
derivative is
\eqn\desir{  
 \delta  (  D_{\mu} \Phi ) = \epsilon * (  D_{\mu} \Phi )  
}
We look for the transformation $ \delta A_{\mu}$
which will guarantee this covariance.
Expanding $\delta  (  D_{\mu} \Phi )$ we get
\eqn\explft{\eqalign{  
& \delta  (  D_{\mu} \Phi ) = \delta ( ( \delta_{\mu} - i A_{\mu} )
* \Phi ) \cr 
& = i \delta_{\mu}  ( \epsilon * \Phi ) - i ( \delta A_{\mu} )* \Phi
+ A_{\mu}* (    \epsilon * \Phi )  \cr
& = i  ( \delta_{\mu}  \epsilon )*\Phi + i  \epsilon *  ( \delta_{\mu}
\Phi ) - i ( \delta A_{\mu} )* \Phi
+ A_{\mu}* (    \epsilon * \Phi ) \cr  }}
In the last line we used the fact the derivatives $ \delta_{\mu} $ 
obey the correct Leibniz rule with respect to the star product
\leibfrst. This manipulation would be more complicated if we used the 
product $ m_2$, as indicated by \copi. 
Expanding the RHS of \desir\ we have 
\eqn\exprht{ 
i \epsilon *(  D_{\mu} \Phi )   =  i \epsilon * ( \delta_{\mu} \Phi
) + 
\epsilon * ( A_{\mu} * \Phi ) 
} 
Equating LHS and RHS we see that we require 
\eqn\delA{\eqalign{  
 ( \delta A_{\mu} ) * \Phi  &=    ( \delta_{\mu}  \epsilon )*\Phi 
- i A_{\mu}* (    \epsilon * \Phi ) + i \epsilon * (   A_{\mu} * \Phi)
\cr 
& =    ( \delta_{\mu}  \epsilon )*\Phi - i (    \epsilon * \Phi )*A_{\mu}  
  + i \epsilon * (   \Phi   * A_{\mu } ) \cr 
}}
In the second line we took advantage of the commutativity 
of the star product to rewrite the RHS. Now the additional 
term in the variation of the gauge field is an associator.
The associator for $3$ general elements 
 $ \Phi_1 , \Phi_2 , \Phi_3$ in $ \cBnst ( R^{2k+1} )$,  
denoted as $ [ \Phi_1 , \Phi_2 , \Phi_3 ] $, is defined as
\eqn\defass{ 
  [ \Phi_1 , \Phi_2 , \Phi_3 ] = 
 ( \Phi_1 *  \Phi_2 ) *  \Phi_3  -  \Phi_1 *  ( \Phi_2   *  \Phi_3 )  
}
We  introduce an operator $ E $ depending on $ \Phi_1, \Phi_3$ 
which acts on $ \Phi_2$ to give the associator. It has an expansion in
powers of derivatives.
\eqn\defCop{\eqalign{  
& [ \Phi_1 , \Phi_2 , \Phi_3 ]  = E( \Phi_1 , \Phi_3) \Phi_2 \cr 
& =  E_{\alpha_1} ( \Phi_1 , \Phi_3)* \delta_{\alpha_1}  \Phi_2 
     +  E_{\alpha_1 \alpha_2 } ( \Phi_1 , \Phi_3) 
* \delta_{\alpha_1}  \delta_{ \alpha_2}   \Phi_2 + \cdots \cr }}
The explicit form of the coefficients in the expansion 
is worked out in Appendix 2 , where we also 
define an operator $ F ( \Phi_1 , \Phi_2) $ related to the 
associator as $ [ \Phi_1 , \Phi_2 , \Phi_3 ] =  F ( \Phi_1 , \Phi_2) \Phi_3 $. 

The gauge transformation of $A_{\mu}$ can now be written as
\eqn\gavar{\eqalign{  
& ( \delta A_{\mu } )*\Phi  = ( \delta_{\mu}  \epsilon )*\Phi  - 
 i  [  \epsilon , \Phi, A_{\mu}   ]   \cr 
&   ( \delta A_{\mu } )   = ( \delta_{\mu}  \epsilon )  -
 i E  ( \epsilon , A_{ \mu } )  \cr 
&  ( \delta A_{\mu } )  = ( \delta_{\mu}  \epsilon )  -
 i  E_{\alpha_1}   ( \epsilon , A_{ \mu } ) * \delta_{\alpha_1} - 
i  E_{\alpha_1 \alpha_2 }   ( \epsilon , A_{ \mu } ) 
* \delta_{\alpha_1}\delta_{\alpha_2} + \cdots  \cr }} 

This leads to a surprise. We cannot in general restrict $ A_{\mu} $ 
to be a function of the $ Z$'s which generate $ \cBnst  ( R^{2k+1}  )$. 
Even if we start with such gauge fields, the gauge transformation 
will cause a change by $  E  (  \epsilon , A_{\mu} ) $ which 
is an operator that acts on $ \cBnst ( R^{2k+1} )$, but is not an operator 
of multiplication by an element of $ \cBnst ( R^{2k+1} )$ as the first 
term is. Instead it involves the derivatives $ \delta_{ \alpha }$.

This means that we should generalize the gauge field $ A_{\mu } $
to $ \hat A_{\mu}$ which  should be understood as having an expansion, 
where $ A_{\mu}$ is just the first term.  
\eqn\genexp{ 
\hat A_{\mu } = A_{\mu} ( Z ) + A_{ \mu  \alpha_1 } ( Z )* \delta_{\alpha_1 } 
 + A_{ \mu  \alpha_1 \alpha_2 } ( Z ) * \delta_{\alpha_1} \delta_
{  \alpha_2  } 
 + \cdots  + A_{ \mu \alpha_1 \cdots \alpha_n } * 
\delta_{\alpha_1} \delta_{  \alpha_2  } \cdots \delta_{\alpha_{n}}
} 
Now if we enlarge the configuration space of 
gauge fields, it is  also natural to enlarge the configuration space 
of gauge parameters, introducing $\hat \epsilon $ 
\eqn\geneps{ 
\hat \epsilon  = \epsilon ( Z ) + \epsilon_{ \alpha_1 } ( Z )*  
\delta_{\alpha_1 } 
+ \epsilon_{ \alpha_1 \alpha_2} ( Z ) * 
\delta_{\alpha_1} \delta_{  \alpha_2  } + \cdots 
+   \epsilon_{\alpha_1 \alpha_2 \cdots \alpha_n } *  
\delta_{\alpha_1} \delta_{  \alpha_2  } \cdots \delta_{\alpha_{n}}    }
This will allow us the possibility that, after 
an appropriate gauge fixing, we recover familiar 
gauge fields which are functions of the coordinates 
only and not coordinates and derivatives. 

On physical grounds, given  that  
the non-associative $2k$-sphere $ \cAn ( S^{2k} )$  
arises as the algebra describing the base space 
of field theory for spherical brane worldvolumes
 \refs{ \clt, \cmt, \cmm,  \sphdiv , \hdim } 
we expect that gauge theory on these algebras, 
and the related algebras $ \cAn^* ( R^{2k+1} ), \cBnst ( R^{2k + 1} ) , \cBnst ( S^{2k } ) $
 should exist. 
We also know that they must approach ordinary 
gauge theory  in the large $n$ limit. 
It is also reasonable to expect that 
the large $n$ limit is approached smoothly. 
Hence we expect that gauge fields which are functions 
of coordinates must be a valid way to describe the 
configuration space at finite $n$. The surprise 
which  comes from exploring the finite $n$ structure of 
the gauge transformations is that such a configuration 
space is only a   partially gauge fixed description.

We write out the gauge transformations of 
$\hat A_{ \mu}$ which follow from covariance of $D_{\mu}$
when we keep the first term in the derivative expansion 
of the generalized gauge transformation $ \hat \epsilon $ 
and the first term in the derivative expansion of
$ \hat A_{\mu} $. The two operators $E$ and $F$ related to 
the associator, which we introduced in appendix 2, are both useful. 
\eqn\ggtrns{ \eqalign{ 
& \delta A_{\mu} = [ \delta_{\mu} \epsilon + i  A_{\mu \alpha_1} 
 * (  \delta_{\alpha_1} \epsilon )   - i \epsilon_{\alpha_1}*  
( \delta_{\alpha_1}  A_{\mu} ) + \cdots  ]   \cr 
& [  - i E_{\beta_1} ( \epsilon , A_{\mu } ) * \delta_{\beta_1} + 
i F_{\beta_1} (  A_{\mu \alpha_1 } , \delta_{\alpha_1}\epsilon ) *  
 \delta_{\beta_1}    - i F_{\beta_1} ( 
 \epsilon_{\alpha_1} , \delta_{\alpha_1} A_{\mu} ) * 
 \delta_{\beta_1} + \cdots 
] \cr 
& + [ -i E_{\beta_1 \beta_2 } ( \epsilon , A_{\mu }  ) * \delta_{\beta_1} 
 \delta_{\beta_2 }  - i E_{\beta_1} ( \epsilon , A_{\mu \alpha_1} ) 
* \delta_{\beta_1} \delta_{\alpha_1} + i F_{\beta_1 \beta_2} ( 
A_{\mu \alpha_1} , \delta_{\alpha_1} \epsilon ) * \delta_{\beta_1}
 \delta_{\beta_2} \cr 
&  - i F_{\beta_1 \beta_2} ( 
\epsilon_{\alpha_1}  , \delta_{\alpha_1} A_{\mu}  ) * \delta_{\beta_1}
 \delta_{\beta_2} + \cdots  ] \cr 
}}
We have separated the terms which involve functions, 
from those involving one derivative operators, and 
two-derivative operators etc. We have not exhibited 
terms involving higher derivative transformations of $A_{\mu} $, 
but it should be possible to exhibit them more explicitly 
using the general formulae for $E$ and $F$  operators 
derived in Appendix 2. If we set the derivative parts 
in $A_{\mu} $ to zero, e.g $ A_{\mu \alpha_1 } =0 $, \ggtrns\ show that 
the purely coordinate dependent pieces still  get modified from the usual 
$ \delta_{\mu } \epsilon $, and from the requirement that the first 
and second derivative corrections to $ A_{\mu  } $ vanish, we get 
\eqn\ggtrnsi{\eqalign{  
&   - i E_{\beta_1} ( \epsilon , A_{\mu } )   
  - i F_{\beta_1} ( 
 \epsilon_{\alpha_1} , \delta_{\alpha_1} A_{\mu} )   
+ \cdots = 0 \cr 
& -i E_{\beta_1 \beta_2} ( \epsilon , A_{\mu} ) -
 i F_{\beta_1 \beta_2 } ( \epsilon_{\alpha_1} , \delta_{\alpha_1} A_{\mu } ) + \cdots = 0 
\cr }}
 For finite $n$ the terms left as dotted should 
be expressible as a finite sum since powers of derivatives higher than 
the $n$'th automatically vanish on any element of the algebra 
$ \cBnst ( R^{2k+1} ) $.

\subsec{ Formula for the associator }

We consider the associator for a triple of general 
elements $ Z_{ \mu ( S )}  ,   Z_{ \mu ( T ) } , Z_{ \mu ( V ) }  $
where $ S, T , V$ are sets of distinct positive integers. 
We use \fzmusmut\ for $ Z_{\mu(S)} * Z_{ \mu ( T) } $. Then 
we compute $  ( Z_{\mu(S)} * Z_{ \mu ( T) } )* Z_{ \mu ( V) }$ as 
\eqn\zmuztznu{\eqalign{  
& ( Z_{\mu(S)} * Z_{ \mu ( T) } )* Z_{ \mu ( V) } 
= \sum_{ |S_1| , |S_2| } \sum_{ |S_3| , |S_4| }
\delta (  |S_1| , |S_2| )  \delta (  |S_3| , |S_4| ) 
\sum_{ S_1 \subset S } \sum_{ S_2 \subset T } \sum_{ S_3 \subset \{ S
\cup T \}  } \sum_{ S_4 \subset V } \cr 
&  { 1 \over n^{2 |S_1| + 2|S_3| } }
{ n! \over ( n - |S_1| ) ! } { n! \over ( n - |S_3| ) ! } 
\tlddel{\mu}{S_1}{S_2} \tlddel{\mu}{ S_3}{S_4} \cr 
&  Z_{\mu ( \{ S \cup T \} \setminus \{ S_1 \cup S_2 \cup S_3 \cup S_4
\} ) } \cr   
}}
It is useful to separate the sum over subsets 
$S_3$ and $S_4$ to distinguish the delta functions 
$ \delta_{ SV } $ between indices belonging to the sets 
$S$ and $V$ and delta functions $ \delta_{ TV } $ between indices belonging to 
the sets $T$ and $V$. We decompose the cardinalities of the sets 
\eqn\sepsums{\eqalign{  
 & |  S_3 |   = | S_{3}^{ (1)}|  + | S_3^{(2)} | \cr 
 & | S_4 | = | S_{4}^{ (1)}|  + | S_4^{(2)} | \cr
}}
and the sets as 
\eqn\sepsets{\eqalign{  
&    S_3 = S_3^{(1)} \cup S_3^{ (2)} \cr 
& \hbox { where} \qquad   S_3^{(1)}  \subset S \setminus S_1 
\qquad S_3^{(2)}  \subset T \setminus S_2  \cr 
&     S_4 = S_4^{(1)} \cup S_4^{ (2)} \cr 
& \hbox { where}   \qquad S_4^{(1)}  \subset  V \qquad 
                  S_4^{(2)}  \subset  V 
}}
The indices in the set $S_3^{(1)}$ are paired with those 
in the set $S_4^{(1)}$ and likewise those in $S_3^{(2)}$
are paired with $S_4^{(2)}$, so that we have $ | S_{3}^{(1)}| = |
S_{4}^{(1)}|$ and $ | S_{3}^{(2)}| = |
S_{4}^{(2)}|$. We express the product in \zmuztznu\ 
\eqn\nzmztzn{\eqalign{ 
& ( Z_{\mu(S)} * Z_{ \mu ( T) } )* Z_{ \mu ( V) } \cr 
& = \sum_{ |S_1| , |S_2| } ~~~~ \sum_{ |S_3^{(1)}| , |S_4^{(1)}| }
~~~~~ \sum_{ |S_3^{(2)}| , |S_4^{(2)}| }
 ~~~~  \delta (  |S_1| , |S_2| ) ~~~ \delta (  |S_3^{(1)}| , |S_4^{(1)}| )
 ~~~ \delta (  |S_3^{(2)}| , |S_4^{(2)}| ) \cr 
& \sum_{ S_1 \subset S  , S_2 \subset T  } ~~~
  \sum_{ S_3^{(1)}  \subset \{ S \setminus S_1 \}  } ~~~
 ~~~~ \sum_{ S_3^{(2)}  \subset \{ T \setminus S_2 \}  }
~~~~ \sum_{ S_4^{(1)}  \subset  V  } ~~~~
 \sum_{ S_4^{(2)}  \subset  V  } ~~ \tlddel{\mu}{S_1}{S_2} ~~ \cr & 
\tlddel{\mu}{S_3^{(1)}}{S_4^{(1)}}~~  
 \tlddel{\mu}{S_3^{(2)}}{S_4^{(2)}}~~  
 Z_{\mu ( \{ S \cup T \} 
\setminus \{ S_1 \cup S_2 \cup S_3 \cup S_4 \} ) } \cr & \qquad 
{ 1 \over n^{2 |S_1| + 2 |S_2 | }} { n! \over ( n - |S_1| ) ! } { n! \over 
( n - |S_3^{(1)} | - |S_3^{(2)} |  ) ! } 
\cr }}
Note that the combinatoric factor needed to rewrite the
 $ \tlddel{\mu}{ S_3}{S_4} $ in terms of the sum of products 
$ \tlddel{\mu}{S_3^{(1)}}{S_4^{(1)}}~~  
 \tlddel{\mu}{S_3^{(2)}}{S_4^{(2)}}$ is $1$. The expression 
$ \tlddel{\mu}{ S_3}{S_4} $ contains $|S_3|!$ Kronecker delta's. 
When we sum the product of two $ \tilde { \delta }$'s we have 
$ { |S_4| \choose |S_4^{(1)}| } = { |S_3| \choose |S_3^{(1)}| }$ as the number 
of ways of picking sets $  S_4^{(1)} $ from $S_4$. When this 
 multiplies the $|S_3^{(1)}|! | S_3^{(2)} |!$  for the number of
Kronecker $\delta$'s 
in the product of two $ \tilde { \delta }$'s we get exactly 
$ |S_3|! $. This explains why the combinatoric factor is $1$.   

In a similar way we can write $ Z_{\mu(S)} * ( Z_{\mu ( T ) } * Z_{\mu
( V)} ) $. After separating the $ \tilde \delta$ from
 the second multiplication 
into a product of two $ \tilde \delta$'s containing pairings of 
 $S$ with $V$ and pairings of $S$ with $T$, we have an expression of 
the same form above except that the factor 
${ n! \over ( n - |S_1| ) ! } { n! \over 
( n - |S_3^{(1)} | - |S_3^{(2)} |  ) ! }$ 
is replaced by 
$ { n! \over ( n - |S_3^{(2)} | ) ! } 
{ n! \over ( n - |S_1| - | S_3^{(1) }) |  ! }$. 
Hence we can write down an expression for the associator 
$ ( Z_{\mu (S)} *  Z_{\mu(T)} ) * Z_{\mu(V)}  - 
  Z_{\mu (S)} *  ( Z_{\mu(T)} * Z_{\mu(V)} )$ as a sum 
of the above form with a coefficient which is a difference of these
two factors. It is instructive to rewrite  the final 
outcome of the multiplications and the associator 
 using a notation for the
summation labels which is more informative and less prejudiced 
 by the order in which the multiplication was done. We write 
 $ R_{ST}$  for the subset of $S$ which is paired with 
 indices in $T$, and $R_{TS} $ for the subset of indices in $T$ 
 which are paired with indices in $S$. The cardinalities are denoted 
 as usual by $ |R_{ST}| = |R_{TS} |$. 
Similarly we use  subsets $R_{SV}$, 
 $ R_{VS}$ for pairings between the sets $S$ and $V$, 
and the subsets $R_{TV}, R_{VT}$. This way we can write 
the associator as 
\eqn\finexpass{\eqalign{  
& ( Z_{\mu (S)} *  Z_{\mu(T)} ) * Z_{\mu(V)}  - 
  Z_{\mu (S)} *  ( Z_{\mu(T)} * Z_{\mu(V)} ) \cr 
& = \sum_{ R_{ST} , R_{TS} } \sum_{ R_{SV} , R_{VS} } \sum_{ R_{TV},
R_{VT} } \tlddel{\mu}{R_{ST}}{R_{TS}}  \tlddel{\mu}{R_{SV}}{R_{VS}} \cr 
& { \tlddel{\mu}{R_{TV}}{R_{VT}} \over 
n^{2 |R_{ST}| + 2 |R_{SV}| + 2 |R_{TV } | }}
 Z_{\mu (~ \{~ S \cup T \cup V \}  \setminus \{ R_{ST} \cup R_{TS} \cup
R_{SV} \cup R_{VS} \cup R_{TV} \cup R_{VT} ~\}~ )}
\cr & 
\bigl (  { n! \over ( n - |R_{ST}| )! } {  n! \over 
( n - |R_{SV}| - |R_{VT}|)! }
 - { n! \over ( n - |R_{TV}| )! } {  n! \over ( n - |R_{ST}| -
|R_{SV}| )! } \bigr ) 
}}
Let us define $A  ( | R_{ST} | , | R_{SV} | , |R_{TV} | ) $ to be the
coefficient in the expression above
\eqn\defA{\eqalign{  
& A  ( | R_{ST} | , | R_{SV} | , |R_{TV} | ) \cr
& \equiv { n! \over ( n - |R_{ST}| )! } {  n! \over 
( n - |R_{SV}| - |R_{VT}|)! }
 - { n! \over ( n - |R_{TV}| )! } {  n! \over ( n - |R_{ST}| -
|R_{SV}| )! }
\cr }}
We observe that exchanging the $S$ and $V$ labels 
in $A$ is an antisymmetry 
\eqn\antsym{ 
A  ( | R_{VT} | , | R_{VS} | , |R_{ST} | ) 
= - A  ( | R_{ST} | , | R_{SV} | , |R_{TV} | ) }
This is to be expected and is nice check on the 
above formulae. Indeed for a { \it commutative } algebra as 
we have here
\eqn\manas{\eqalign{  
& [Z_{\mu(S)} , Z_{\mu(T)} , Z_{\mu(V)} ] 
= ( Z_{\mu (S)} *  Z_{\mu(T)} ) * Z_{\mu(V)}  - 
  Z_{\mu (S)} *  ( Z_{\mu(T)} * Z_{\mu(V)} ) \cr 
& =   Z_{\mu (V)} *  ( Z_{\mu(S)} * Z_{\mu(T)} )  - 
   ( Z_{\mu(T)} * Z_{\mu(V)} ) *Z_{\mu (S)} \cr 
&=  Z_{\mu (V)} *  ( Z_{\mu(T)} * Z_{\mu(S)} ) 
   - ( Z_{\mu(V)} * Z_{\mu(T)} ) *Z_{\mu (S)}\cr 
& = -  [Z_{\mu(V)} , Z_{\mu(T)} , Z_{\mu(S)} ] \cr }}

\subsec{ Simple Examples of Associator } 

As another check on the general formula 
\finexpass\ we consider an example of the form 
$ [ Z_{\mu(S)} , Z_{\mu_1} , Z_{\mu_2} ] $. 
\eqn\frstexm{\eqalign{  
& (  Z_{\mu(S)} *  Z_{\mu_1} ) = Z_{\mu_1 \mu(S) }  + 
\sum_{ i\in S } { 1 \over n } 
 \delta_{\mu_1 \mu_i} Z_{ \mu ( S \setminus i ) } \cr 
& (  Z_{\mu(S)} *  Z_{\mu_1} )*Z_{\mu_2} 
 =  Z_{\mu_1 \mu_2 \mu(S) }  + { \delta_{\mu_1 \mu_2} \over n }
Z_{\mu(S) } + \sum_{i \in S } { 1 \over n } 
 \delta_{\mu_2 \mu_i} Z_{ \mu_1 \mu ( S \setminus i ) }\cr 
& + \sum_{ i, j \in S } { 1 \over n^2 } 
 \delta_{\mu_1 \mu_i}\delta_{\mu_2 \mu_j} Z_{ \mu ( S \setminus i ,j  ) }
\cr 
} }

When we multiply in the other order 
\eqn\secex{\eqalign{  
& Z_{\mu_1} * Z_{\mu_2} = Z_{\mu_1 \mu_2} + { \delta_{\mu_1 \mu_2}
\over n } \cr 
&   Z_{\mu(S)} *  ( Z_{\mu_1} *Z_{\mu_2} ) 
= Z_{\mu_1 \mu_2 \mu(S) } + { 1 \over n } \sum_{ i \in S }
\delta_{\mu_1 \mu_i} Z_{\mu_2 \mu  ( S \setminus i )} +  
{ 1 \over n } \sum_{ i \in S }
\delta_{\mu_2 \mu_i} Z_{\mu_1 \mu ( S \setminus i )} \cr & 
+ { n(n-1) \over n^4} \sum_{ i , j \in S }
\delta_{\mu_1 \mu_i} \delta_{\mu_2 \mu_j }  Z_{ \mu ( S \setminus i ,
j ) } +   { \delta_{\mu_1 \mu_2} \over n }
Z_{\mu(S) } \cr }}
Hence the associator is 
\eqn\assexf{\eqalign{ 
& [ Z_{\mu(S)} , \mu_1 , \mu_2 ] = 
( Z_{\mu(S)} * Z_{\mu_1 } ) * Z_{\mu_2 } -  
Z_{\mu(S)} * ( Z_{\mu_1 }  * Z_{\mu_2 } ) \cr 
&= { 1 \over n^3 } \sum_{ i , j \in S }
\delta_{\mu_1 \mu_i} \delta_{\mu_2 \mu_j }  Z_{ \mu ( S \setminus i ,
j ) }}}
Note that if $|S| =1 $ the associator is zero. 
The single $ \delta$ terms cancel in the associator. 
This follows from \finexpass\ because $ A ( 1, 0, 0 ) = A( 0,1,0 ) =
A( 0,0,1) = 0 $. The coefficient $ {1 \over n^3}$ of the two $\delta$ terms 
can be read off from \finexpass\ by noting 
that for $ |R_{ST}| = |R_{SV}| = 1 $ and $ |R_{TV}| =0$ , 
we have 
\eqn\coefnt{\eqalign{  
& {1 \over n^{2 |R_{ST}| + 2 |R_{SV}| + 2 | R_{TV} |}} A ( |R_{ST}| ,
|R_{SV}| , |R_{TV}| ) 
\cr & = { 1 \over n^4 } [  n^2 - n(n-1) ]  =
{ 1\over n^3
} \cr }}
The expressions developed above can lead to the construction of operators
$E$ and $F$ related to the associator. The operators are described in Appendix 2 
and used in the first part of this section. 

\newsec{ Gauge Theory on $ \cBnst ( S^{2k} )$ }

We will obtain an action for the deformed algebra of functions 
$ \cBnst ( S^{2k} )$ which is obtained by applying 
the constraint of constant radius to the algebra 
$ \cBnst ( R^{2k+1} )$.

\subsec{ A finite $n$ action which approaches the abelian
          Yang-Mills action on $S^{2k}$ } 
 
 We begin by writing the ordinary  pure Yang Mills 
 action on $S^{2k}$ in terms of the embedding 
 coordinates in $R^{2k+1}$. We describe the $S^{2k}$ in terms of $
 \sum_{\mu }  z_{\mu} z_{\mu}  =
 1 $. The derivatives $  { \partial \over \partial z_{\mu } } $ 
 can be expanded into an angular part and a radial part. 
\eqn\expnsdmu{\eqalign{  
\dr{\mu} &= { \partial \theta^{a} \over \partial z_{\mu} } \prtder
 { \theta^a } + { \partial r \over \partial z_{\mu} } \prtder{r} \cr 
& = P_{\mu} +  { z_{ \mu} \over r } \prtder{r} \cr }}
We have denoted as $P_{\mu} = P_{\mu}^a  \prtder { \theta^a }   $ 
the projection 
of the derivative along the sphere. Consider the commutator 
$   [ P_{\mu}  - i A_{\mu} ,  P_{\nu}  - i
 A_{\nu}] $ for gauge fields $A_{\mu} $ satisfying $ z_{\mu} A_{\mu} =
 0 $ and which are functions of the angular variables only. A general
 gauge field $A_{\mu}$ has an expansion $ A_{\mu} =  P_{\mu}^a A_a +
 { z_{\mu} \over r } A_r $. The condition $ z_{\mu} A_{\mu} =0 $ 
 guarantees that $A_r = 0 $, hence 
\eqn\gfang{ 
 A_{\mu} = P_{\mu}^a A_{a} } 
The following are useful observations
\eqn\stps{\eqalign{ 
&[  D_{\mu} ,  D_{\nu } ] =  [ P_{\mu}  -i A_{\mu} , P_{\nu}  -i A_{\nu} ] \cr
&  =   P_{\mu}^a P_{\nu}^b F_{ab} + L_{\mu \nu }^a D_a  \cr 
& =  P_{\mu}^a P_{\nu}^b F_{ab} +  z_{\mu} D_{\nu} - z_{\nu} D_{\mu} \cr 
 }}
We can then express 
\eqn\rexp{ 
P_{\mu}^a P_{\nu}^b F_{ab} = [  D_{\mu} ,  D_{\nu } ] - 
( z_{\mu} D_{\nu} - z_{\nu} D_{\mu} ) 
} 
Now observe that the $P_{\mu}^a P_{\mu}^b = G^{ab}$, i.e the 
inverse of the metric induced on the sphere by its embedding 
in $R^{2k+1}$, which is also the standard round metric. 
Hence we can write the Yang Mills action as 
\eqn\ymact{ 
 \int d^4 \theta { \sqrt G }  G^{ac} G^{bd} F_{ac}F_{bd} 
   = \int d^4 \theta { \sqrt G }  ( [ D_{\mu} , D_{\nu} ]  - 
       ( z_{\mu} D_{\nu} - z_{\nu} D_{\mu} ) )^2 
}

We define a commutative non-associative $ \cBnst ( S^{2k+1} ) $ 
by imposing conditions 
$ Z_{\alpha \alpha } =  { ( n-1 ) \over n }R^2   $.
$Z$ operators with more indices contracted 
e.g $ Z_{\alpha \alpha \beta \beta }$ can be deduced 
because they appear in  products like  
$ Z_{\alpha \alpha} * Z_{ \beta \beta }$
and their  generalizations. We have introduced an extra 
generator $R$ which corresponds to the radial coordinate 
transverse to the sphere.

The projection operators $P_{\alpha}$ obey 
$ P_{\alpha } = \delta_{\alpha } - Q_{\alpha} $. 
We have already shown that $ \delta_{\alpha}$ acts as 
a derivation of the star product. The behaviour of $ Q_{\alpha} $ 
can be obtained similarly. In general we can expect it to be a deformed 
derivation.   
In classical geometry : 
\eqn\zalph{\eqalign{  
& \delta_{\alpha} R = { Z_{\alpha} \over R }  \cr 
& Q_{\alpha} R = { Z_{\alpha} \over R }  \cr 
& Q_{\alpha} Z_{\beta} = { Z_{\alpha} Z_{\beta} \over R^2 } \cr 
}}  
This suggests that the definitions at finite $n$ : 
\eqn\zalphi{\eqalign{  
& \delta_{\alpha} R = { Z_{\alpha} \over R }  \cr 
& Q_{\alpha} R  = { Z_{\alpha} \over R }  \cr 
& Q_{\alpha} Z_{\mu ( S) } = { |S| \over R^2 }  Z_{\alpha} * Z_{ \mu(S) } \cr  
}}
In the classical case $Q_{\alpha}$ is a derivation 
so its action on a polynomial in the $Z$'s is defined 
by knowing its action on the $Z$'s and using Leibniz rule. 
In the finite $n$ case we have given the action 
on a general element of the algebra and we leave 
the deformation of the Leibniz rule to be computed. 
The deformed Leibniz rule can be derived by techniques similar 
to the ones used in section $3$. Following section $5$, we can determine 
the gauge transformation of $A_{\mu}$ by requiring covariance 
of $D_{\mu} = P_{\mu } - i A_{\mu}$. The deformation 
of Leibniz rule implies that the gauge transformation law 
will pick up extra terms due to 
\eqn\extrtrms{\eqalign{  
  P_{\alpha} ( \epsilon * \Phi )& =
 ( \delta_{\alpha} - Q_{\alpha} ) ( \epsilon * \Phi ) \cr 
& = ( \delta_{\alpha} \epsilon )* ( \Phi ) + 
\epsilon*(  \delta_{\alpha} \Phi )   - ( Q_{\alpha} \epsilon )* \Phi 
- \epsilon* ( Q_{\alpha} \Phi ) - (Q_{\alpha}^{(1)} \epsilon) * 
(Q_{\alpha}^{(2)} \Phi) + \cdots 
\cr 
}}
The last term is a schematic indication of the 
form we may expect for the corrections to the 
Leibniz rule, following section 2. 
More details on the deformed Leibniz rule
for $Q_{\alpha }$ are given in Appendix 3.

As in section 5, the non-zero associator will 
require generalizing the gauge parameters and 
gauge fields in order to make the gauge invariance manifest. 
It is not hard to write down a finite $n$ 
action ( without manifest gauge invariance ) 
 which reproduces the classical action on the sphere
\eqn\finact{\eqalign{ 
 \int  (  [ P_{\mu} - i A_{\mu } , P_{\nu } - i A_{\nu} ]   - 
          ( Z_{\mu } * D_{\nu } - Z_{\nu } * D_{ \mu } )  ]^2  
}}
It has the same form as \ymact. 
The $A_{\mu } $ are chosen to satisfy $ Z_{\mu} * A_{\mu} = 0 $ 
to guarantee they have components tangential to the sphere, 
and they are expanded in symmetric traceless polynomials 
in the algebra $ \cBnst ( S^{2k} )$. The integral is defined 
to pick out the coefficient of the integrand which is proportional to 
the identity function in the algebra $ \cBn ( S^{2k} )$. 
It is reasonable to guess that this action can be obtained 
as a gauge fixed form of an action where the gauge invariances 
are manifest at finite $n$. The equations to be solved now 
involve the operators $E$ and $F$ as well as the additional 
terms due to the corrections to Leibniz rule for $Q_{\alpha}$.

\newsec{ Outlook  }

\subsec{ The non-abelian generalization and Matrix Brane constructions  } 
In the bulk of this paper we have considered 
abelian theories on a class of commutative 
but non-associative algebras related to 
the fuzzy $2k$-spheres. It is also possible to generalize the 
calculations to non-abelian theories. Indeed these are the kinds 
of theories that show up in brane constructions { \clt , \cmt , \cmm }. 
In this case   
we need to  consider covariant derivatives of
the form $ \delta_{\mu} -i A_{\mu} $ where the 
$A_{\mu}$ are matrices whose entries take 
values in the $ \cAn ( S^{2k} ) $. The $ \Phi $ are also 
Matrices containing entries which take values in the 
algebra.  Again we define the transformation of 
$ \delta \Phi = \epsilon * \Phi $ or writing out to include the 
Matrix indices  $ \delta \Phi_{ij}  = \epsilon_{ik} * \Phi_{kj} $. 
As in the abelian case, we can construct the gauge transformation law 
for $A_{\mu}$ by requiring covariance of $ P_{\mu } - i A_{\mu } $. 
It will  very interesting 
to construct  these non-abelian generalizations in detail. 
They would answer a very natural question which arises in the 
D-brane applications where one constructs a higher brane worldvolume 
from a lower brane worldvolume. There is evidence from 
D-brane physics that the fuzzy sphere ansatze on the lower brane worldvolume 
leads to physics equivalent to that described by the higher 
brane worldvolume with a background non-abelian field strength. 
Following the logic of \aikkt\ it should be possible to 
start with the lower brane worldvolume and derive an action 
for the fluctuations which is a non-abelian theory on the 
higher dimensional sphere. Thus the evidence gathered for the agreement 
of the physical picture ( brane charges, energies etc. ) 
 from lower brane and higher brane worldvolumes 
would be extended into a derivation of the higher brane action from 
the lower brane action. Since the only algebra structure 
we can put on the truncated spectrum of fuzzy spherical harmonics 
at finite $n$ is a non-associative one \sphdiv, 
the  techniques of this paper provide  an avenue for constructing 
this non-abelian theory.

An alternative description of the fluctuations in these 
Matrix brane constructions
 is in terms of an abelian theory on a 
higher dimensional geometry $ SO(2k+1)/U(k)$ \refs{  \hdim, \kimura}. 
Combining this with the physical picture discussed above 
of a non-abelian theory on the commutative/non-associative 
sphere, we are lead to expect that there should be a duality 
between the abelian theory on the  $ SO(2k+1)/U(k)$ and 
the abelian theory on $S^{2k}$ \hdim.  The explicit construction of 
the detailed form of the non-abelian theories on the 
the fuzzy sphere algebras $ \cAn ( S^{2k} ) $ will also be interesting 
since it will allow us to explore the mechanisms of this duality. 
There has been a use of effective field theory 
on the higher dimensional geometry in the quantum hall effect 
\refs{ \zhhu, \bchtz }.

\subsec{ Physical applications of the abelian theory } 

 In applications of the fuzzy $4$ sphere 
 to the ADS/CFT as in \jr\berkverl\ 
 we have abelian gauge fields in spacetime ( in addition to gravity 
fields etc. ) 
 so we should expect an abelian gauge theory 
 on the fuzzy sphere if we want the fuzzy 
 4-sphere to be a model for spacetime. 
 In the context of the  
 of longitudinal $5$-branes \clt\ in the DLCQ ( Discrete 
 Light cone quantization ) of M-Theory, 
 it was originally somewhat surprising 
 that only spherical $5$-brane numbers of the form 
 $  { ( n +1 ) ( n + 2 ) ( n+ 3 ) \over 6 }  $ for integer  
 $n$ could be described. We may ask for example 
 why there is not a DLCQ description of a single spherical 
 $5$-brane.  The abelian field theories 
 considered here would be candidate worldvolume 
 descriptions  for the single $5$-brane in the  $DLCQ$ of M-theory, 
 but it is not clear how to directly derive them from 
 BFSS Matrix theory.

\subsec{ Other short comments }

We have only begun to investigate the 
structure of gauge theories related to 
commutative, non-associative algebras. 
For applications to Matrix Theory fuzzy 
spheres, it will be interesting to 
work directly with $ \cAn ( R^{2k+1} ) $ 
and $ \cAn ( S^{2k} )$ ( along with  the twisted algebras 
$ \cAn^* ( R^{2k+1} ) $ and $ \cAn^* ( S^{2k} )$ ), instead 
of the $ \cBn (  R^{2k+1} ) $ and $ \cBn ( S^{2k} )$
 ( along with  their twisted versions $ \cBnst (  R^{2k+1} ) $
  $ \cBnst  ( S^{2k} )$ ) which were introduced for simplicity, 
by considering a large $D$ limit of  $ \cAn ( R^{D+1 } ) $.  
Instanton equations for gauge theories on 
these algebras will be interesting. 
In connecting these field theories on $ \cAn ( S^{2k} ) $ 
to the Matrix Theory fuzzy $2k$-spheres, it will be necessary 
to expand around a spherically symmetric  instanton background. 
Developing gauge fixing procedures and setting up 
perturbation theory are other directions to be explored. 
The projection operators $ P_{\alpha}$ together 
with the $SO(2k+1) $ generators $ L_{ \mu \nu } $ 
form the conformal algebra $ SO(2k+1,1)$. It will 
be interesting to see how a finite $n$ deformation 
of this acts on the finite $n$ commutative/non-associative 
algebra of the fuzzy $2k$-sphere.

By replacing $ \delta_{\mu \nu }$ with 
$ \eta_{\mu \nu }$ ( the $SO(2k,1)$ Lorentzian invariant ) 
in \simpnass\strcprod \fzmusmut\  
we get  commutative, non-associative algebras which 
deform Lorentzian spacetime. Since the commutator 
$ [ Z_{\mu } , Z_{\nu } ] $ is zero, we do not 
have a $ \theta_{\mu \nu } $ which breaks Lorentz invariance. 
If we assign standard dimensions of 
length to $Z$ and introduce a dimensionful deformation parameter $\theta $ 
to make sure all terms in \strcprod\fzmusmut\ are of same 
dimension, we only need a $ \theta$ which is $scalar$ under the Lorentz group. 
It will be very interesting to study these algebras as 
Lorentz invariant deformations of spacetime in the context of 
the transplanckian problem for example, along the lines of 
\branho.

The generalized gauge fields and gauge parameters 
which we were lead to, as a consequence of formulating 
gauge theory for the non-associative algebras, are strongly 
reminiscent of constructions that have appeared in 
descriptions of the geometry of $W$-algebras \hullw, 
where higher symmetric tensor fields appear as gauge parameters 
and gauge fields. 
It is possible that insights from W-geometry can be 
used in further studies of  gauge theories on 
commutative, non-associative algebras. 
If a connection to W-algebras can be made, the equations 
such as \ggtrnsi\ and its generalizations, 
constraining the generalized gauge parameters  \geneps\ 
by requiring that the gauge fields remain purely coordinate 
dependent, would be conditions which define an embedding of 
the $U(1) $ gauge symmetry inside a W-algebra.  The embedding 
is trivial when the non-associativity is zero, given simply 
by setting the $ \epsilon_{\alpha_1} , \epsilon_{\alpha_1 \alpha_2}  
\cdots  = 0  $, but requires the higher tensors to be non-zero and related 
to each other when the non-associativity is turned on. 
It is also notable that generalized gauge fields involving dependence on derivatives 
have appeared  \pmnass\ when considering non-associative algebras related to
non-zero background $H$ fields \cornschiap.  
Further geometrical understanding of these features 
 appears likely to involve the BV formalism \batvil.

\bigskip
\noindent{\bf Acknowledgments:}
 I would like to thank Luigi Cantini, David Gross, Bruno Harris, 
 Pei Ming Ho, Chris Hull, David Lowe, Shahn Majid,
 Robert de Mello-Koch , Rob Myers,  Martin Rocek,  Jae Suk Park, 
 Bill Spence,  Dennis Sullivan, Gabriele Travaglini 
 for very instructive  discussions. 
 I thank the organizers of the Simons Workshop at Stony Brook
 for hospitality  while  part of this work was done.
 This research  was supported  by DOE grant  
 DE-FG02/19ER40688-(Task A) at Brown, and 
 by a PPARC Advanced Fellowship  at Queen Mary. 

\bigskip 
\bigskip 

\vfill\eject 

\noindent 
{ \bf Appendix 1 : The classical product in terms of the star product $m_2^*$ 
 }

 We introduce 
 an algebra  $ \cBnc ( R^{2k+1}  ) $ with product $m_2^c $ 
 which, as a vector space, is the same as $ \cBn ( R^{2k+1}  ) $
 and $  \cBnst ( R^{2k+1}  ) $. The product resembles the 
 classical product in that it is just concatenates the 
 indices on the $Z$'s. 
\eqn\concat{ 
m_2^c ( Z_{\mu ( S) } \otimes Z_{ \mu ( T ) } ) 
=  Z_{\mu ( S \cup  T ) } 
}  
The  formula  \fzmusmut\ for the product 
 $m_2^*$ can be rewritten in terms of 
$m_2^c$ 
\eqn\mtwostcon{ 
m_2^* = \sum_{ k=0 }^{n} { 1 \over n^{2k}  k! }    { n! \over (n- k)! }
 \sum_{ \alpha_1 \cdots \alpha_k } ~~ m_2^c ~.~  ( \delta_{\alpha_1} \cdots 
\delta_{ \alpha_k }    
\otimes   \delta_{\alpha_1} \cdots \delta_{ \alpha_k }    )          
} 
It is very useful to have the inverse formula 
where $m_2^c$ is expressed in terms of $m_2^*$. 
This can be used to write general formulae for the associator
by manipulating \finexpass. 
\eqn\mtwoconst{ 
m_2^c = \sum_{ k=0 }^{n} { (-1)^k \over n^{2k-1}  k! }    { ( n +k-1 )! 
 \over n! }  
 \sum_{ \alpha_1 \cdots \alpha_k }  ~~ m_2^* ~.~ ( \delta_{\alpha_1} \cdots 
\delta_{ \alpha_k }    
\otimes   \delta_{\alpha_1} \cdots \delta_{ \alpha_k }  )            
} 
Proving this inversion formula makes use of a combinatoric identity.
Let us define $f_{a} ( n)  = { n! \over (n-a)! }  { 1 \over n^{2a} } $, 
and ${ \tilde f}_{ a} ( n) = 
{(-1)^a  \over n^{2a-1} }  { ( n+a-1)! \over n! } $. 
The combinatoric identity is 
\eqn\Ffcomb{\eqalign{  
% & 
{\tilde f}_k ( n ) = 
\sum_{l=1}^{k} (-1)^l \sum_{  a_1, a_2,\cdots a_l \ge 1 : \sum a_i = k    }
  f_{a_1}f_{a_2 } 
\cdots f_{a_l}  
%\cr  &  \qquad \qquad \qquad \qquad 
{ k! \over a_1! a_2! \cdots a_l! } . { l! \over | S(a_1, a_2,
\cdots a_l )| }\cr }} 
We are summing over ways of writing the positive integer 
$k$ as a sum of integers $a_i \ge 1  $.  For each such splitting 
of $k$, the symmetry factor $ S( a_1, a_2, \cdots a_l )$ 
is the product $ n_1 ! n_2 ! \cdots $ where $ n_1$ is the number 
of $1$'s , $n_2$ the number of $2$'s etc. We have checked the 
identity for $k$ up to $6$ using Maple and find the simple form 
of $\tilde f$ given above. An analytic proof for general $k$ 
would be desirable.

\bigskip

\noindent 
{ \bf Appendix 2 : Operators related to the Associator }

In \finexpass\ we have written down the associator 
of three general elements of the algebra $ \cBnst ( R^{2k+1}) $ 
in terms of $ Z_{ \mu (~ \{~ S \cup T \cup V \}  
\setminus \{ R_{ST} \cup R_{TS} \cup R_{SV} \cup R_{VS} \cup 
R_{TV} \cup R_{VT} ~\}~ )} $. This can be expressed in terms of the 
product $m_2^c$ 
\eqn\expmtwc{\eqalign{  
& Z_{\mu (~ \{~ S \cup T \cup V \}  
\setminus \{ R_{ST} \cup R_{TS} \cup R_{SV} \cup R_{VS} \cup 
R_{TV} \cup R_{VT} ~\}~ )} \cr 
& = m_2^c . ( m_2^c \otimes 1) . 
  (  Z_{ \mu ( S \setminus \{ R_{ST} \cup R_{SV} \} )} \otimes 
   Z_{ \mu ( T \setminus \{ R_{TS} \cup R_{TV} \}) }  \otimes 
   Z_{ \mu ( V \setminus \{ R_{VS} \cup R_{VT} ) }   ) \cr 
& =  m_2^c . ( m_2^c \otimes 1) .
 (  Z_{ \mu ( S \setminus \{ R_{ST} \cup R_{SV} \} )} \otimes 
   Z_{ \mu ( V \setminus \{ R_{VS} \cup R_{VT} \} ) }    \otimes 
  Z_{ \mu ( T \setminus \{ R_{TS} \cup R_{TV} \}) }   ) \cr 
}}
By using \finexpass\ together with \expmtwc\
and converting $m_2^c$ to $m_2^*$ we can 
get formulas for the associator in terms of $m_2^*$. We 
\eqn\assoc{\eqalign{  
& [ \Phi_1, \Phi_2 , \Phi_3 ] 
= \sum_{r_1 =0 }^{n}  \sum_{r_2 =0 }^{n}  \sum_{r_3 =0 }^{n} 
\sum_{ a_1 \cdots a_{r_1} } \sum_{ b_1 \cdots b_{r_2} }
\sum_{ c_1 \cdots c_{r_3} } \cr 
& \sum_{m=0}^{n} \sum_{\beta_1 \cdots \beta_m} 
\sum_{l=0}^{n} \sum_{\alpha_1 \cdots \alpha_l} 
{ {\tilde f }_{l} ( n ) \over l! } { {\tilde f}_{m} ( n ) \over m! } 
{ 1 \over r_1! r_2!r_3! } [ f_{r_1} (n) f_{r_2+r_3} (n) - f_{r_3} (n) 
f_{r_1+r_2} (n) ] \cr 
& \delta_{ \beta_1 \cdots \beta_m } ( \delta_{\alpha_1} \cdots \delta_{\alpha_{l}} 
 \delta_{a_1} \cdots \delta_{a_{r_1} } \delta_{b_1} \cdots \delta_{ b_{r_2} }
 \Phi_{1} * \delta_{\alpha_1} \cdots \alpha_{l} 
\delta_{a_1} \cdots \delta_{a_{r_1} } \delta_{c_1} \cdots \delta_{ c_{r_3} }
 \Phi_2  ) * \cr & 
\delta_{ \beta_1} \cdots \delta_{ \beta_{m} } 
\delta_{b_1} \cdots \delta_{b_{r_2} } \delta_{c_1} \cdots \delta_{ c_{r_3} } 
 \Phi_{3} 
\cr  }}
The $a,b,c, \alpha , \beta $ indices all run from $1$ to $2k+1$. 
This expresses the associator $[ \Phi_1 , \Phi_2 , \Phi_3 ] $ 
as an operator $ F $ depending on $ \Phi_1 , \Phi_2 $ and acting on 
$ \Phi_{3} $ 
\eqn\oponthr{ 
[ \Phi_1 , \Phi_2 , \Phi_3 ] = { F}_{\alpha_1 } ( \Phi_1 , \Phi_2 )
* \delta_{\alpha_1} 
 \Phi_3  + {F}_{ \alpha_1 \alpha_2} ( \Phi_1 , \Phi_2 )
* \delta_{\alpha_1} \delta_{\alpha_2 }  \Phi_3 + \cdots }
The successive terms are subleading in the $1/n$ expansion. 
 
We can also use the second line of \assoc\ to  write the associator as 
\eqn\assoi{\eqalign{ 
& [ \Phi_1 , \Phi_2 , \Phi_3 ] 
=  
 \sum_{r_1 =0 }^{n}  \sum_{r_2 =0 }^{n}  \sum_{r_3 =0 }^{n} 
\sum_{ a_1 \cdots a_{r_1} } \sum_{ b_1 \cdots b_{r_2} }
\sum_{ c_1 \cdots c_{r_3} } \cr 
& \sum_{m=0}^{n} \sum_{\beta_1 \cdots \beta_m} 
\sum_{l=0}^{n} \sum_{\alpha_1 \cdots \alpha_l} 
{ {\tilde f}_{l} ( n ) \over l! } { {\tilde f }_{m} ( n ) \over m! } 
{ 1 \over r_1! r_2!r_3! } [ f_{r_1} (n) f_{r_2+r_3} (n) - f_{r_3} (n) 
f_{r_1+r_2} (n) ] \cr 
& \delta_{ \beta_1 \cdots \beta_m } ( \delta_{\alpha_1} \cdots \alpha_{l} 
 \delta_{a_1} \cdots \delta_{a_{r_1} } \delta_{b_1} \cdots 
\delta_{ b_{r_2} } \Phi_{1} *
\delta_{\alpha_1} \cdots \alpha_{l}\delta_{b_1} \cdots  \delta_{ b_{r_2} }
\delta_{c_1} \cdots \delta_{ c_{r_3} } 
 \Phi_{3} ) * \cr & 
\delta_{\beta_1} \cdots \beta_{m} \delta_{b_1} 
\cdots \delta_{b_{r_1} } \delta_{a_1} \cdots \delta_{ a_{r_1} }
 \Phi_2   \cr  }}
This gives the associator as an operator $E$ depending 
on $ \Phi_1, \Phi_3 $ and acting on $ \Phi_2$
\eqn\opontw{ 
[ \Phi_1 , \Phi_2 , \Phi_3 ] = E_{\alpha_1 } ( \Phi_1 , \Phi_3 )  
* \delta_{\alpha_1}  \Phi_3  + 
E_{ \alpha_1 \alpha_2 } ( \Phi_1 , \Phi_3 ) * \delta_{\alpha_1} 
\delta_{\alpha_2 }  \Phi_2 + \cdots }
 The successive terms are subleading in the $1/n$ expansion.

\bigskip 

\vfill\eject 

\noindent 
{ \bf Appendix 3 : Deformed Leibniz Rule for $Q_{\alpha} $      }

Using the definition of $Q_{\alpha} $  we can work out
the deformation of its Leibniz rule. 
\eqn\defleibq{\eqalign{  
& Q_{\alpha} ( Z_{\mu( S) } * Z_{ \mu( T ) } ) - 
( Q_{\alpha}  Z_{\mu( S) } )  * Z_{ \mu(T) } 
-  Z_{\mu( S) }*  ( Q_{\alpha} Z_{ \mu(T) } )  \cr  
& = - { 1 \over R^2} \sum_{ T_1 \subset S } \sum_{ T_2 \subset T }
\tlddel{\mu}{T_1}{T_2} { 2 |T_1| \over n^{2|T_1| } }
      { n! \over ( n- |T_1| )! }
 Z_{\alpha \mu ( S \cup  T \setminus \{ T_1 \cup T_2 \} )}  \cr 
& + { 1 \over R^2 } \sum_{T_1 \subset S } ~~~~ \sum_{ i , T_2 \subset T } 
    ( { -2 |T_1| \over n } + { |T_1|  |S| \over n } ) { n! \over n^{2 |T_1| }
 ( n - |T_1| ) ! } \cr 
& \qquad \qquad \qquad \qquad \delta_{\alpha \mu_i } \tlddel{\mu}{T_1}{T_2} 
Z_{ \mu ( S \cup T \setminus \{ T_1 \cup T_2 \} ) }
       \cr & 
+ { 1 \over R^2} \sum_{i, T_1 \subset S } ~~~~~ \sum_{ T_2 \subset T } 
( { -2 |T_1| \over n } + { |T_1|  |T| \over n } ) 
{ n! \over n^{2 |T_1| } ( n - |T_1| ) ! }
\cr & \qquad \qquad \qquad \qquad  
\delta_{\alpha \mu_i } \tlddel{\mu}{T_1}{T_2} Z_{ \mu ( S \cup T \setminus 
        \{ T_1 \cup T_2 \} ) } }}
We can write this deformation of the Leibniz rule in terms of 
operators acting on $ Z_{\mu ( S) } \otimes Z_{\mu(T) } $ as follows  
\eqn\expdefqmc{\eqalign{  
& Q_{\alpha} ( Z_{\mu( S) } * Z_{ \mu( T ) } ) - 
( Q_{\alpha}  Z_{\mu( S) } )  * Z_{ \mu(T) } 
-  Z_{\mu( S) }*  ( Q_{\alpha} Z_{ \mu(T) } )  \cr 
& = - { 1 \over R^2} \sum_{ k=0}^{n} \sum_{a_1 \cdots a_k} 
 { 2 k \over n^{2k} } { n! \over  k!  ( n- k ) ! } 
   m_2^c ( m_2^c \otimes 1 ) \cr 
&       
      ( 1 \otimes \delta_{a_1} \cdots \delta_{a_k} 
                  \otimes \delta_{a_1} \cdots \delta_{a_k} ) 
           ( Z_{\alpha}  \otimes Z_{ \mu ( S ) } \otimes Z_{ \mu ( T) } ) \cr 
& +{ 1 \over R^2} \sum_{k=0}^{n} \sum_{ b_1 \cdots b_k} 
{ 1 \over n^{2k}}  { n! \over k! ( n- k ) ! } 
                      ( { - 2 k \over n } + { |S| k \over n^2 } ) \cr 
&                       m_2^c ( m_2^c \otimes 1 )  
( 1 \otimes \delta_{b_1}\cdots \delta_{b_k} \otimes   
 \delta_{b_1}\cdots \delta_{b_k} ) ( \delta_a \otimes 1 \otimes \delta_a ) 
 ( Z_{\alpha} \otimes Z_{ \mu(S) } \otimes Z_{ \mu( T) } ) \cr 
& + { 1 \over R^2} \sum_{ k=0 }^{n } \sum_{ b_1 \cdots b_k}
{ 1 \over n^{2k}}  { n! \over k! ( n- k ) ! } 
( { - 2 k \over n } + { |T| k \over n^2 } )\cr  
& m_2^c ( m_2^c \otimes 1 )  
   ( 1 \otimes \delta_{b_1} \cdots \delta_{b_k} \otimes 
\delta_{b_1} \cdots \delta_{b_k}   ) 
( \delta_{a} \otimes \delta_{a} \otimes 1 ) ( Z_{\alpha} \otimes Z_{\mu(S) } 
         \otimes Z_{ \mu(T) } )  }}
Now we can convert the classical  products 
$m_2^c$ into expressions involving $m_2^*$ 
using the first Appendix. 
\eqn\expdefqmst{\eqalign{
& Q_{\alpha} ( Z_{\mu( S) } * Z_{ \mu( T ) } ) - 
( Q_{\alpha}  Z_{\mu( S) } )  * Z_{ \mu(T) } 
-  Z_{\mu( S) }*  ( Q_{\alpha} Z_{ \mu(T) } )  \cr 
& =  - { 1 \over R^2} \sum_{ k=0}^{n} \sum_{a_1 \cdots a_l}
\sum_{l=0}^{n} \sum_{l=0}^{n} \sum_{b_1 \cdots b_l =1  }^{2k+1} 
 { 2 k \over n^{2k} } { n! \over  k!  ( n- k ) ! } 
{ \tilde f_l (n) \over l! }  \cr & 
  m_2^{*} \qquad \qquad ( \delta_{c_1} \cdots \delta_{c_l} \otimes
 \delta_{c_1} \cdots \delta_{c_l} )  \qquad \qquad 
[ ~~~ ( m_2^* \otimes 1 ) + \tilde f_1(n) ( m_2^* \otimes 1 )
( \delta_{b_1} \otimes \delta_{b_1} \otimes 1 )~~ ]  \cr & 
  ( 1 \otimes \delta_{a_1} \cdots \delta_{a_k} 
 \otimes \delta_{a_1} \cdots \delta_{a_k}  )    
           ( Z_{\alpha}  \otimes Z_{ \mu ( S ) } \otimes Z_{ \mu ( T) } ) \cr
&             
+ { 1 \over R^2} \sum_{k=0}^{n} \sum_{b_1 \cdots b_k } 
\sum_{ l=0}^n \sum_{c_1 \cdots c_l} 
 { 1 \over n^{2k}}  { n! \over k! ( n- k ) ! } 
                      ( { - 2 k \over n } + { |S| k \over n^2 } ) 
{ \tilde f_l (n) \over l! } \cr 
& 
                      m_2^* \qquad \qquad 
( \delta_{c_1} \cdots \delta_{c_l} \otimes \delta_{c_1} \cdots \delta_{c_l} ) 
 \qquad \qquad    ( m_2^* \otimes 1 )  \cr & 
( 1 \otimes \delta_{b_1}\cdots \delta_{b_k} \otimes   
 \delta_{b_1}\cdots \delta_{b_k} ) ( \delta_a \otimes 1 \otimes \delta_a ) 
 ( Z_{\alpha} \otimes Z_{ \mu(S) } \otimes Z_{ \mu( T) } ) \cr 
& 
+ { 1 \over R^2} \sum_{ k=0 }^{n } 
\sum_{b_1 \cdots b_k }  \sum_{ l=0}^n \sum_{c_1 \cdots c_l} 
{ 1 \over n^{2k}}  { n! \over k! ( n- k ) ! }
        ( { - 2 k \over n } + { |T| k \over n^2 } ) { \tilde f_l (n) \over l! }
          \cr &             m_2^*  \qquad \qquad 
( \delta_{c_1} \cdots \delta_{c_l} \otimes \delta_{c_1} \cdots \delta_{c_l} ) 
  \qquad \qquad   ( m_2^* \otimes 1 )  \cr &   
( 1 \otimes \delta_{b_1} \cdots \delta_{b_k} \otimes 
\delta_{b_1} \cdots \delta_{b_k}   ) 
( \delta_{a} \otimes \delta_{a} \otimes 1 ) ( Z_{\alpha} \otimes Z_{\mu(S) } 
         \otimes Z_{ \mu(T) } )  }}
Since an arbitrary element of $ \cBnst ( R^{2k+1 } ) $ can be expanded 
in terms of $ Z_{\mu ( S) } $ we can replace 
$  Z_{\mu ( S) } $  and $ Z_{\mu ( T) } $ above by 
arbitrary elements $ \Phi_1 , \Phi_2$ of the algebra. 

\bigskip

\noindent 
{\bf Appendix 4 : $m_2^c $ in terms of $m_2$  } 

  We have found above that expressing the  
 non-associativity of $m_2^*$ in terms of 
 operators and  finding the deformed 
 Leibniz rule for $Q_{\alpha}$ with respect to 
 $m_2^*$ are conveniently done by having 
 a formula for the classical product in terms 
 of $m_2^*$. In this paper we started with 
 $m_2$ and found that if we twist it into 
 $m_2^*$ we get another commutative, non-associative 
product with nicer properties, notably one on which 
 $\delta_{\alpha}$ acts as a derivation. 
 If one chooses to work with $m_2$, it is convenient 
 to have a formula for $m_2^c$ in terms of 
 $m_2$. Let us first note that the defining formula 
for $m_2$ \strcprod\ can be expressed as a formula 
 for $m_2$ in terms of $m_2^c $
\eqn\rwtcrc{ 
 m_2 = \sum_{k=0}^{n} \sum_{a_1 \cdots a_k } m_2^c . 
( \delta_{a_1} \cdots \delta_{a_k} \otimes \delta_{a_1} \cdots \delta_{a_k} ) 
 { ( n - D + 2k ) ! \over k! ( n-D +k )! }
}
$ D $ acting on an element of $ \cBn \otimes \cBn $ is 
just the sum of the degrees of the two elements. For 
 example on $ Z_{ \mu ( S) } \otimes Z_{ \mu ( T) } $ it is 
$ |S| + |T| $. Let us define $ g_{k} ( n , D ) = 
{ ( n - D + 2k ) ! \over k! ( n-D +k )! } $
Inverting \rwtcrc\ to give $m_2^c$ in terms of $ m_2^c$ 
to find the coefficients $ \tilde g_{k} ( n, D ) $ which appear in the 
inverse formula
\eqn\rwtcrc{ 
 m_2^c  = \sum_{k=0}^{n} \sum_{a_1 \cdots a_k } m_2. 
( \delta_{a_1} \cdots \delta_{a_k} \otimes \delta_{a_1} \cdots \delta_{a_k} ) 
\tilde g_{k} ( n, D ) }
leads to sums of the form 
\eqn\sumfrm{ 
\tilde g_k ( n, D ) = \sum_{ l=0}^n \sum_{a_1 \cdots a_l} 
g_{a_1} ( n, D ) g_{a_2}( n , D-2a_1  ) \cdots g_{a_l} ( n, D - 2a_1 -2a_2 
\cdots  2a_{l-1} ) { k! \over a_1! a_2! \cdots a_l! } 
} 
There is a sum over $l$, and for each $l$ a  sum 
 over choices of ordered sets of 
positive integers $ a_1 , a_2 \cdots a_l $ which satisfy 
$ a_1 + a_2 + \cdots a_l = k $. 
The first few examples of $ \tilde g_1, \tilde g_2, \tilde g_3, \tilde g_4,
 \tilde g_5 $ 
were done by using Maple ( some of them are also easily checked by hand ). 
They all agree with  a simple general formula which is 
\eqn\Gk{ 
 \tilde g_{k} ( n, D ) = { (-1)^k \over n^2k} ( n - D + 2k ) 
{ ( n - D + k -1 ) ! \over ( n - D ) ! } 
} 
An analytic proof for general $k$ 
would be desirable. This $\tilde g_k$ can be used to 
write the associator and deformed Leibniz rules 
involving the product $m_2$.

\listrefs

\end